\documentclass[fleqn,10pt]{wlscirep}
\usepackage[utf8]{inputenc}
\usepackage[T1]{fontenc}
\usepackage{upgreek}
\usepackage{color}
\definecolor{darkblue}{rgb}{0,0,0.5}
\definecolor{lila}{rgb}{0.3,0,0.3}
\definecolor{turq}{rgb}{0,0.1,0.4}
\definecolor{lightblue}{rgb}{0.7,0.7,0.9}
\usepackage{url} 

\newcommand{\done}{\ensuremath{\mathbf{1}}}
\newcommand{\dzero}{\ensuremath{\mathbf{0}}}

\title{Single Photon Randomness based on a Defect Center in Diamond}

\author[1]{Xing Chen}
\author[1]{Johannes Greiner}
\author[1,2]{J\"org Wrachtrup}
\author[1,*]{Ilja Gerhardt}
\affil[1]{3. Institute of Physics, University of Stuttgart and Institute for Quantum Science and Technology (IQ$^{\mbox{ST}}$), Pfaffenwaldring 57, D-70569 Stuttgart, Germany}
\affil[2]{Max Planck Institute for Solid State Research, Heisenbergstra\ss e 1, D-70569 Stuttgart, Germany}

\affil[*]{i.gerhardt@pi3.uni-stuttgart.de}

\begin{abstract}
The prototype of a quantum random number generator is a single photon which impinges onto a beam splitter and is then detected by single photon detectors at one of the two output paths. Prior to detection, the photon is in a quantum mechanical superposition state of the two possible outcomes with --ideally-- equal amplitudes until its position is determined by measurement. When the two output modes are observed by a single photon detector, the generated clicks can be interpreted as ones and zeros -- and a raw random bit stream is obtained. Here we implement such a random bit generator based on single photons from a defect center in diamond. We investigate the single photon emission of the defect center by an anti-bunching measurement. This certifies the ``quantumness'' of the supplied photonic input state, while the random ``decision'' is still based on the vacuum fluctuations at the open port of the beam-splitter. Technical limitations, such as intensity fluctuations, mechanical drift, and bias are discussed. A number of ways to suppress such unwanted effects, and an a priori entropy estimation are presented. The single photon nature allows for a characterization of the non-classicality of the source, and allows to determine a background fraction. Due to the NV-center's superior stability and optical properties, we can operate the generator under ambient conditions around the clock. We present a true 24/7 operation of the implemented random bit generator.
\end{abstract}
\begin{document}

\flushbottom
\maketitle
%
%
\thispagestyle{empty}

\section{Introduction}

The generation of random bits is a growing field of quantum science. A primary quantum process and a subsequent measurement allow to implement an inherently unpredictable measurement outcome. In this sense a quantum randomness generator follows Born's rule, which describes the measurement outcome of a quantum measurement as fundamentally probabilistic. Further, the interest in quantum random numbers is testified by the fact that the most popular available products in quantum technology are quantum random number generators. Some companies are selling quantum random number generators, which emit a stream of ones and zeros, supposed to fulfill the crucial requirements of a random number: The number of ones and zeros is approximately equal, and no memory and no correlation among two or more bit outcomes is evident -- the numbers are \emph{independent and identically distributed}. Most commonly a user has problems to verify that the emitted stream is based on a true quantum process and not deterministically pre-programmed or at least predictable by an (evil) manufacturer or other adversary. The latter would be feasible since a recent discussion in cryptography identified the supply of \emph{weak} random numbers as a reasonable attack vector of modern cryptography~\cite{lenstra__2012,becker__2013}. Such \emph{kleptographic} attacks~\cite{young_cr_1996} pose an important threat to modern information security technology. Therefore the question exists how such devices might allow a user to examine and to ensure the integrity of a primary random process and follow the randomness processing.

Apart from early, mostly theoretical descriptions~\cite{schmidt_joap_1970,erber_n_1985,rarity_jomo_1994}, a variety of quantum random number generators were implemented around the turn of the century~\cite{stefanov_jmo_2000,jennewein_rosi_2000}. These early generators were based on the measurement of photonic qubits. With the increasing demand of speed and the implementation of coherent states, a number of generators were implemented which use other light source than single photons. These generators often measure the vacuum fluctuations~\cite{trifonov_patent_2007,shi_apl_2016,steinle_prx_2017} or phase noise~\cite{xu_oe_2012,abellan_oe_2014}. In the past years the implementation of device independent randomness generators~\cite{colbeck_phd_2009,pironio_n_2010} or self-testing schemes~\cite{lunghi_prl_2015} has been experimentally demonstrated. A timely review has been written~\cite{herrero-collantes_rmp_2017}.

Although a single photon which impinges on a beam splitter and is later detected is still the most prominent model system for a quantum decision, only very few implementations of true \emph{single} photon randomness have been demonstrated -- mostly due to the implied technical complications and the questionable benefit of such an implementation. The few experiments introduce single photons from a down-conversion source which impinge on a beam splitter~\cite{bronner_ejop_2009,branning_josab_2010}, or on an integrated optics device~\cite{graefe_nphot_2014}.

As for single photon sources, a large variety of single emitters as single photon sources has been investigated in the past. One prominent example is the negatively charged nitrogen-vacancy center. It has been experimentally singled out since 1997~\cite{gruber_s_1997}. Besides its properties as nanoscopic sensor and tool for spin based quantum information processing, it represents a stable single photon source with up to few million counts per second~\cite{kurtsiefer_prl_2000,brouri_ol_2000,jamali_rosi_2014}. A variety of single photon based implementations were realized based on defect centers in diamond, for example quantum cryptography~\cite{beveratos_prl_2002}, or other fundamental experiments~\cite{jacques_s_2007}. Only few experiments were performed which utilize the single photon emission of a NV-center for quantum randomness generation. Only recently a single NV-center in a sample was utilized for quantum randomness generation based on the polarization detection of a single photon~\cite{abe_srep_2017}.

Here we theoretically discuss and experimentally implement a single photon based quantum random number generator. The single photon stream originates from a single nitrogen-vacancy-center in diamond. The device is operated for more than a week in continuous operation. The entropy of the raw bit stream is estimated with a conservative a priori model. The single photon nature allows for a characterization of the non-classicality of the source, but yields naturally a lower randomness rate than generators, which rely on the detection laser noise or laser phase fluctuations~\cite{abellan_oe_2014}. Furthermore the amount of unwanted, technical background clicks can be estimated by this model. In how far this leads to a certifiable randomness generation is discussed. The advantage of a true single photon source allows a user to exclude a number of attack scenarios and to guarantee to a certain extend the independence to an external adversary of the device.

\section{Theory}

Under discussion is a random bit generator, in which the outcomes on two discrete single photon detectors in different output modes of a beam-splitter are interpreted as single raw bits. This section describes the fundamental ideas to process the raw detector outcomes, i.e.\ the electrical ``clicks'' of the two detectors. These bits then go into some post-processing, namely a randomness extraction process. For the calculation of the underlying entropy, the outcome probabilities and the transition probabilities are relevant. These single and transition probabilities are calculated in full in the supplementary material. In total three models are introduced, which describe increasingly more conservative interpretations of the raw-bits.

\subsection{Single Photon Stream}

A stream of single photons is guided in a spatially single optical mode towards a beam-splitter. The second input arm of this beam-splitter is blocked, which is commonly described as a vacuum state ($|0\rangle$) in the second input arm. Therefore, the single photons probability amplitudes are distributed on the beam-splitter according to the beam-splitter ratio. The photons are subsequently detected on the detectors $A$ (transmitted photons) and $B$ (reflected photons). 

For this first model, we simply consider \emph{all} generated clicks as raw random bits. The probability of the individual outcomes and the transition probabilities are relevant. Here we discuss the fundamental parameters to estimate the entropy in the next section. Our interpretation is not limited to the use of a single photon input state, but it might still be advantageous due to an increased click rate on a single photon detector by the interplay of detector dead times and the characteristic time constant of a single emitter~\cite{oberreiter_light_2016}.

The probability whether a photon is detected in one or the other output path is fundamentally linked to the presence of the vacuum state at the second input port of the utilized beam-splitter. Further, it is dependent on a variety of experimental factors. Most importantly, the beam splitter ratio $\mathcal{R}$ is a function of the reflection ($R$) and transmission ($T$) coefficient. For simplicity, a loss-less beam-splitter is assumed, i.e.\ $T+R=1$. This does not imply an equality of the two coefficients, i.e.\ the beam-splitter can still be biased. Although a biased beam splitter does introduce some imbalance, it does not necessarily introduce any memory in the system, and can therefore represent a perfect Bernoulli trial with a given probability distribution.

Another crucial parameter in the (im-)balance of the detector recordings is the detector efficiency of the utilized detectors, $\eta_{\mathrm{\{A,B\}}}$. This value describes how many of the incident photons actually lead to an electrical pulse which can then be recorded. In a given experimental implementation, the individual detector efficiency is closely linked to the beam-splitter ratio. Both factors can not be independently measured in a simple way in a given setting.

An electrical pulse from a single photon detector has to have a finite length, such that it can be recorded with normal detection hardware. During this time, a second detection event is usually suppressed. This time of suppression is called the ``dead-time'', $\tau_{\mathrm{dead}}$, of the detector. In the common Geiger mode photo detector modules the time where no second photon is detected, commonly exceeds the length of the electrical pulse. This dead-time can be described by a correction factor, which relates the click rate to the actual single photon flux onto the detector. At low incident photon counts and low dark-counts of the detector, the factor is close to unity. This factor accounts for the undetected photon events, which are suppressed as a second event which should have been launched although the detector was still found within its dead-time.

The technicality of the detector dead times also introduces another problem for the generation of random bits: Two subsequent clicks on different detectors can be dependent, since one of the detector is in its dead time, only the other detector is capable of detecting another photon. In the limiting case for very high count-rates this leads to an alternating bit pattern with zero entropy, because when the detector is able to detect a next photon, it immediately receives one and generates another click. This leads to an anti-correlation of detector events, and becomes dominant at higher count rates where the probability of coincidental clicks is non-zero~\cite{oberreiter_light_2016}.

Subsequently, the total click rate on the detectors is $r_{\mathrm{total}}=r_{\mathrm{A}}+r_{\mathrm{B}}$. It amounts to the following expression. To recall: the behavior between generated raw bits and the incident intensity is not strictly linear.

\begin{equation}
r_{\mathrm{A,B}} = \eta_{\mathrm{A,B}}T I_{\mathrm{in}}-\frac{(\eta_{\mathrm{A,B}}T I_{\mathrm{in}})^2\int_0^{\tau_{\mathrm{dead}}^{\mathrm{A,B}}}g^{(2)}_{}(\tau)d\tau}{4}
\end{equation}
where $\eta_{\mathrm{A,B}}T I_{\mathrm{in}}$ is the click rate when no dead-time would be present, and $g^{(2)}(\tau)$ is the anti-bunching curve~\cite{photon_anti_re} of the utilized single photon source.

This equation accounts for the click rates independently of the input light source. A laser source obeys an exponential decaying probability distribution in the subsequent detection of a photon. At the same incident photon flux, a single photon source has a higher probability of a later detection event and might generate a larger randomness generation rate than a laser with the same brightness. This relates to an interplay of the detector's dead-time and the count rates. This is outlined in more detail outlined in~\cite{oberreiter_light_2016}.

The detection rates for the two utilized detectors have been determined by the consideration of the beam splitter ratio, the individual detection efficiency and the individual dead times. The \emph{probabilities} for calculating the entropy below can be experimentally determined straight forwardly by the ratio of the detector events:

\begin{equation}
p_{\mathrm{A}} = \frac{r_{\mathrm{A}}}{r_{\mathrm{A}}+r_{\mathrm{B}}}
\label{eqn:single}
\end{equation}

\noindent
The conditional probabilities are correspondingly calculated as follows. Here we exemplary show the derivation for the case of the conditional detection of $p(A|A)$:

\begin{equation}
p(A|A)=\big(1-\int_0^{\tau^{\mathrm{A}}_{\mathrm{dead}}}g^{(2)}(\tau)d\tau\big)\eta_{\mathrm{A}}T
\end{equation}

\noindent
This is outlined in the supplementary material in more detail. The deviation of the conditional probabilities from the ideal value of 0.5 is dominated by the combination of the detector efficiencies and the beam-splitter ratio. The specific detector dead times influence the conditional probabilities of the conditional photon detection events only at increased count rates.

In this section we have considered that \emph{all} generated photon detection events act as a source of raw quantum randomness. The underlying entropy relates to the bias and other technical consideration such as the detector efficiency and the dead-time of the detectors. The calculated probabilities and the conditional probabilities are utilized to calculate the min-entropy below.


\subsection{Bounds on Quantumness}

The above description of photon click rates is applicable correspondingly to a simple laser or an light-emitting diode (LED) input source. Equivalently, we would effectively sample the vacuum fluctuations which determine the output of the measurement at the beam-splitter. On the other hand, the use of a laser has some drawbacks: When a coherent state, $|\alpha\rangle$, is present, the user has no indication if the device \emph{really} detects the incoming (laser) mode. A counter example could be that the device was misaligned in the past and only uncorrelated photons from two independent sources are detected. The two beams which impinge from a laser via the beam-splitter onto the detectors are unrelated in a number of ways, and cannot be differentiated if two different lasers (or two independent laser modes) were observed with the detectors. In the worst case, an eavesdropper will remote-control the detector clicks and due to their initially uncorrelated origin, the owner of the device will have no proof on their origin or integrity.

Therefore, another strategy of an eavesdropper could be to simply send uncorrelated counts into the two detectors. In the case of the laser source, this flaw would not be indicated by any measurable quantity, except that possibly more counts per unit time are present.

A single photon source is of advantage at this point and might lead to some enhancement of the trust in the random bit generation scheme. As an ultimate proof to have a single photon emitter, an auto-correlation function can be recorded. For a single photon source, this shows the typical single photon anti-bunching. This gives an upper bound on how much randomness can be extracted, when we assume that the random bit generator is equipped with a single photon emitter. The amount of (uncorrelated) counts, which might be remotely controlled events, can be well determined. Furthermore, a certain guarantee is given, that one single mode is detected and the detector does not observe e.g.\ uncorrelated sources.

Naturally, for a single photon input state, the photons arrive at the beam-splitter in an \emph{anti-bunched} fashion. This implies, that after a photon detection event on one detector, the second detector does not observe a single click within a characteristic time.

Anti-bunching is commonly described by the auto-correlation function $g^{(2)}(\tau)$. In an ideal case $g^{(2)}$ for zero time delay ($\tau=0$) amounts to zero, and at a photon detection event no other detection is observed. In this case, all the detected raw bits are originating from a fundamental quantum process, and we consider all raw bits as quantum random bits. In reality, a recording of $g^{(2)}(0)=0$ is very hard to achieve: The dark counts of the detectors, any kind of background contributions from the environment, and also the electrical jitter of the experimental devices, are factors which will be mixed with the single photon events and increase the final value of $g^{(2)}(0)$. Since all these spurious background contributions introduce uncorrelated events, we consider such events as being based on clicks which can be known to an external adversary called ``Eve''. Subsequently, this fraction of uncorrelated events is discarded by a later randomness extraction process.

The term \emph{true single photons} denotes all the photons which are detected by the detectors and stem from the device-internal single photon source. Usually, measurement artefacts reduce this fraction from unity. Still this fraction can be well determined by the so-called anti-bunching curve, as $\sqrt{1-g^{(2)}(0)}$. The anti-bunching characterizes the quantum nature of the involved photons, and it cannot be interpreted by classical theory~\cite{photon_anti_re}. Subsequently, the single photons which are characterized by anti-bunching are generated by a fundamental quantum process, while their distribution over the detectors (still) relies on the vacuum fluctuations on the beam splitter.

Also, the recording of the anti-bunching curve allows to estimate a background fraction. This is performed by analyzing the correlation function at zero time delay. This allows to give an estimation on the ``quantumness'' of the single photon stream. While in the model above \emph{all} photons were considered, we now bind the amount of randomness generation clicks on the amount of true single photons in the stream. This requires the assumption, that the randomness generator contains a single photon source and that the underlying randomness is bound on the vacuum state on the second input port of the beam splitter.

In order to estimate the amount of click events which are generated by a quantum process, the fraction of single photon events in the entire photon stream is determined. The fraction of single photon events is known as $\sqrt{1-g^{(2)}(0)}$~\cite{Brouri:00}, where the non-zero value of $g^{(2)}(0)$ testifies a background fraction. Subsequently, all the generated raw-clicks can be reduced by hashing down to the conservative bound of the below calculated entropy. When $g^{(2)}(0) \geq 1$, all the raw random data will be discarded. In this case, the source for the generator is no longer based on the device internal single photon source~\cite{photon_anti_re,loudon2000quantum}, and the bit generation process is likely to be externally influenced and in the worst case completely controlled by Eve.

For the calculation of the probabilities, we refer to the discussion above. The consideration of background events implies that a few of the generated bits are actually known to an eavesdropper. Therefore it is not clear if they were generated by the genuine device, or if they were introduced in an uncorrelated manner. This fraction is accounted for in the supplementary material by introducing a probability, that the eavesdropper is aware of the generated bit, $p_e$. This discussion extends also to the derivation of the conditional probabilities, which are required to estimate the min-entropy of the generator under these considerations.

\subsection{Conditioned Tuple Detection of Detector Clicks}

As outlined above, every recording of anti-bunched light is subject to background influences. This usually originates from the sample under study or external sources such as ambient light. In the worst case --especially under conservative considerations-- this might also originate from an external adversary, who controls the detectors in an uncorrelated fashion. Therefore, all considerations on the entropy of this single photon stream were bound to the amount of ``true'' single photons, which can be estimated by the recording of anti-bunching. Such a recording exhibits a small fraction of photons, which are time-wise anti-correlated -- the anti-bunching dip. On the other hand, an auto-correlation function around $g^{(2)}(\tau)$=1 implies that the photonic emitter has no memory, and also no photon-photon correlation, for a particular waiting time $\tau$. This is given for photon detection events which are time-wise separated, and do not have a ``partner-photon'' which can be used to characterize the (anti-)correlation. Above we have assumed, that the recording of anti-bunching describes the \emph{full} data stream, even if a photon was detected with no further closely neighboring detection events, which would then have been suppressed and would directly testify the anti-bunching.

This assumption is based on the idea that the photonic stream is fair sampled. This means that the photons which are not contributing to the small time-window of the anti-bunching dip (approx.\ 1-30~ns, corresponding to the T$_1$-time of the system) are assumed to be still genuine single photons. Of course, statistical evidence allows us to consider this assumption is proven. Ideally, a normalized anti-bunching dip is below unity, but some parameters of the experimental devices may have fluctuations, and may cause that the anti-bunching exceeds the value of unity. To guarantee the quantumness of the extracted bits from the raw data, we now only consider the data in a time window below the line of unity (classical line). By considering the fluctuation of the single photon events, we limit the area of random bits to the area which is clearly below this classical boundary. This implies a time-wise selection to events which have a ``partner-photon'' temporally close-by.

As for the recording of the anti-bunching curve, only the \emph{tuples} of photon detection events are considered. These are events which are often described as ``start-stop events'' on two photon detectors behind a beam splitter. Fundamentally, these tuples are balanced, since the probability of detecting the bit pair ``AB'' and ``BA'' depends on the individual probabilities of detecting photons and -- the corresponding pair-detection probabilities are equal $p$(AB)=$p$(BA), as long as the experimental parameters do not change in the course of the experiment. While this tuple detection scheme introduces an advantage due to its inherently balanced nature, it also reduces drastically the detection rate of the raw randomness events, since only start-stop events in a limited time range are utilized. This start-stop event detection scales approximately quadratically (more insights below) with the single count-rates and is usually orders of magnitude smaller than in the simple methods which are described above.

For the selection of anti-bunched detection tuples, we only select the bits which are clearly below a level of $g^{(2)}(\tau)$=1. These events are clearly anti-correlated, but are still above the background level in the $g^{(2)}(\tau)$-recording. In this area below the anti-bunching curve, we consider the pair ``AB'' as one random bit, ``BA'' as the other random bit (Note here that we are dealing with binary random bits, so we only have two random bits: 0 or 1.). These are events, where one detector clicked, while the other one had just fired. This implies that these newly defined raw bits are bound to a detection of a pair of photons in both arms.

Ideally, a normalized anti-bunching curve of an ideal two-level system exists only below unity. This implies, that the considered values \emph{all} fulfill the non-classical nature of $g^{(2)}(\tau)<$1. In a real-world scenario, some parameters of the experimental devices may have fluctuations. These may be introduced as shot noise. To guarantee the quantum nature of the utilized light source, in this situation, we consider the data below unity with given standard deviations, for example, $11.5\sigma$ deviations, which limits the probability of an outlier significantly.

An external adversary has now limited options to influence the device based on the strategy of tuple detection: Any uncorrelated event which is controlled from an external entity will lead to an increased background fraction. This would also be suppressed by the prior method on limiting the amount of randomness to the single photon nature of the entire stream and discarding the background. Therefore, an external adversary would have to implement more subtle strategies to launch clicks in the generator.

A more subtle strategy could be, that when a click is launched by the primary process, i.e.\ an emitted photon from the single photon source, Eve has to quickly launch a click onto the other detector of the generator -- simply since the first detector will be in its dead time. As discussed below, this requires a stringent timing of the clicks launched by Eve. First, Eve has to \emph{detect} that there has been a click in the generator; then she has to \emph{introduce another click} into the generator. If the legitimate time frame is very short, Eve has to be very close to the generator -- simply due to the signal traveling time. Obviously, if Eve is outside the light cone, she has no option to launch clicks and to influence the generator at her will. In this sense, the generated bits which are based on short time differences (-ns) in subsequent clicks in the generator are ``fresh'' and guaranteed to be unaltered for a short amount of time.

Another, more sophisticated attack version of Eve can be to fake the clicks on both detectors, as if Eve had another single photon source at her location. Then, the number of clicks, which are outside the small nanosecond range of the primary source is large. Depending on the relative brightness compared to the primary source this will lead to a certain amount of background clicks again. The only option that Eve has, is to fully suppress the primary emission of the single photon source and replay an equivalent detector control scheme which would introduce a comparable anti-bunching signal. A simple control of the primary photon source by another method can reveal that an external adversary is active.

To calculate the probabilities for the next entropy extraction step, the start-stop event (tuple) probabilities $p(AB)$ and $p(BA)$ are considered. Furthermore, the conditional probabilities (e.g.\ $p(AB|AB)$) have to be described. For a full derivation see the supplementary material.

\section{A Priori Entropy Analysis and Randomness Extraction}

We now turn to the amount of extractable entropy for the generator. This is the key measure for any randomness generator and allows to post-process the raw input bits. Naturally every generator has to perform some post processing, since the true amount of randomness in any experimental configuration can only be determined to a certain level of accuracy. Then one has to consider, that even a sophisticated characterization run is just an accidental outcome. Subsequently, we have to consider a conservative bound, which allows to reduce the number of bits to the true entropy fraction which is available from the raw random stream. This reduces the amount of bits which are extracted from the raw bits.

For all the three discussed cases above, we need to estimate the conditional min-entropy for post-processing the random bits. The definition of the required conditional min-entropy, $H_\infty$ is given as~\cite{simplebounds,oberreiter_light_2016}:

\begin{equation}
\begin{split}
H_{\infty}(X|Y)=-\mathrm{log}_2 \Big(\sum_y p(y) \max\limits_{x}\{p(x|y)\}\Big)\, ,
\label{eqn:condmin}
\end{split}
\end{equation}

\noindent
where $x$ and $y$ are the subsequent events of random bits. In this definition the single and the transition probabilities are the crucial parameters, and they can all be derived from the experimental parameters as discussed above.

For the first case described above, we want to use two universal hashing as an extractor to generate a nearly uniformly distributed random string~\cite{renner_a_2004}. In the generated raw random bits, the number of zeros and ones are not always the same. Different parameters may influence this bias between zeros and ones. For the security of the random bits, the conditional min-entropy is calculated. The conditional min-entropy is defined as before (Eqn.~\ref{eqn:condmin}).

The conditional min-entropy is a more strict bound than conditional Shannon entropy. With the theoretically estimated bound of $H_{\infty}(X|Y)$, randomness extraction could be applied to achieve a nearly uniformly distributed random string. To be specific, from the experimental parameters, the value of $H_{\infty}(X|Y)$ could be derived by the relevant probabilities, suppose $H_{\infty}(X|Y)=k$, then for the $n$ bits long raw random data, approximately $k n$ bits nearly uniform distributed random data could be extracted. The generation speed of the raw bits is $r_{\mathrm{total}}$, then the generation speed of the unbiased random bits amounts accordingly to $k r_{\mathrm{total}}$.

For the second case, only the single photon fraction of the assumed fair sampled stream is considered. We now estimate the min-entropy of the stream under consideration of the background fraction. This requires that our model is extended with a third possible outcome, which is associated to an eavesdropper or classical background light. The outcomes would therefore be $A$, $B$ and the events known to an eavesdropper $E$. The probabilities are not only considered as genuine outcomes of the two detectors, but a third probability is introduced, which denotes that the generator's outcome was not generated by the given single photon source, but by Eve. This fraction reduces the overall amount of entropy of the generator. The probability of a \emph{known} outcome, which generated by Eve, is a function of the observed $g^{(2)}_{\mathrm{fit}}(0)$, which indicates any deviation to a true single photon stream.

As shown in the supplementary material, the fraction of pure single photon events is $\sqrt{1-g^{(2)}_{\mathrm{fit}}(0)}$. The probability of having an event from the eavesdropper relates accordingly to the $g^{(2)}$-function: $p_e=1-\sqrt{1-g^{(2)}_{\mathrm{fit}}(0)}$. This fraction of non-single photon events may also change the conditional probability of $p(x|y)$.

Now, all input parameters for the generator are defined. We discard probabilities in our consideration which could be known by an eavesdropper. Also events, which are influenced by an eavesdropper are neglected. This reduces the equation to:

\begin{equation}
\label{eqn:e_conmin}
\begin{split}
&H_{\infty}(X|Y)\\
&=-\mathrm{log}_2 \Big(p_e+(1-p_e)\big(\sum_y p(y) \max\limits_{x}\{p(x|y)\}\big)\Big)\, .
\end{split}
\end{equation}

\noindent
Compared to the above Eqn.~(\ref{eqn:condmin}), the extractable entropy given by the Eqn.~(\ref{eqn:e_conmin}) is reduced. However, the random bits which can be extracted by this equation are generated by single photon events, which means they are considered to originate from the genuine source of the generator. Suppose $H_{\infty}(X|Y)=k_{\mathrm{q}}$ in this case, then with an $n$ bits long raw data stream, accordingly $k_{\mathrm{q}}n$ quantum bits could be extracted from the raw bits. The generation speed of this unbiased quantum random number corresponds to $k_{\mathrm{q}}r_{\mathrm{total}}$.

For the last case, only paired events and the \emph{area} below unit line (i.e.\ $g^{(2)}_{\mathrm{fit}}(\tau)<=1$) are considered. This line is treated as the classical limit~\cite{photon_anti_re,loudon2000quantum,fox_book_2006}. This area is used to determine the fraction of very conservative considered single photon quantum randomness from the raw data. In order to estimate this fraction, the click rates of the raw events are required. For the convenience of description, this area is named as the ``quantum area'' in the following.

The quantum randomness fraction can be derived by considering the start-stop events for a given timing resolution $\tau_{\mathrm{rs}}$. The start-stop event rate, $r_{\mathrm{stsp}}$, of the two detectors and uncorrelated events (e.g.\ laser emission) is given for a certain time resolution $\tau_{\mathrm{rs}}$ as:

\begin{equation}\nonumber
r_{\mathrm{stsp}}=r_{\mathrm{A}}\times r_{\mathrm{B}}\times \tau_{\mathrm{rs}}\, .
\end{equation}

We like to note the difference by a factor of two against our prior work~\cite{oberreiter_light_2016}, which is caused by disregarding the order of the events in this case.

\noindent
In the ``quantum area'', at different delay times, the start-stop events in a given timing resolution, $\tau_{\mathrm{rs}}$, correspond to different $g^{(2)}(\tau)$ values. This means they obey different probabilities~\cite{fox_book_2006}. The experimental anti-bunching curve is represented as $g^{(2)}_{\mathrm{fit}}(\tau)$. Then, the total photon start-stop event rate in this quantum area is given by

\begin{equation}\nonumber
\begin{split}
&r_{\mathrm{A}}\times r_{\mathrm{B}}\times \sum_{\tau=-t}^{\tau=t}\tau_{\mathrm{rs}}g^{(2)}_{\mathrm{fit}}(\tau)\approx\\
&r_{\mathrm{A}}\times r_{\mathrm{B}}\times \int_{-t}^{t} g^{(2)}_{\mathrm{fit}}(\tau)d\tau\, ,
\end{split}
\end{equation}

\noindent
where $t$ satisfies $g^{(2)}_{\mathrm{fit}}(t)=1$, which means that the entire range is considered until the events are not anti-correlated any more. $g^{(2)}_{\mathrm{fit}}(\tau)$ is the anti-bunching curve with background, which means the above equation also includes the start-stop events caused (partially) by background noise. In order to deal with this situation, we use, as above, the fraction $\sqrt{1-g^{(2)}_{\mathrm{fit}}(0)}$ to exclude the background noise in the clicks of each detector. Then the start-stop events originating from the single photon events are

\begin{equation}
r_{\mathrm{stsp}}=(1-g^{(2)}_{\mathrm{fit}}(0))\times r_{\mathrm{A}}\times r_{\mathrm{B}}\times \int_{-t}^{t} g^{(2)}_{\mathrm{fit}}(\tau)d\tau\, .
\label{rand}
\end{equation}

\noindent
Their count rate in this area is the generation speed of single photon events which are short-time related. The tight bound of the randomness generation to the genuine source of the described generator is guaranteed by the short time distance of the start-stop events $AB$ and $BA$. Therefore, the generation speed of the quantum random bits in this part is linked to the start-stop events as $r_{\mathrm{rand}}=r_{\mathrm{stsp}}$.


With this start-stop event count rate, the fraction of extractable quantumness per raw bit is determined.

Notice that the quantum random bits which are generated tight to this area are supposed to be well balanced. The raw bits are generated at a speed of $r_{\mathrm{total}}$, so the fraction of quantum random bits from per raw bit is $r_{\mathrm{rand}}/r_{\mathrm{total}}$. This value is affected by the shape of the anti-bunching curve and $g^{(2)}_{\mathrm{fit}}(0)$. An extreme case is that when the background noise dominates, such as $g^{(2)}_{\mathrm{fit}}(0)$ is unity. In this case, we can not extract any quantum randomness from the raw bit data.

Since the fraction of quantum randomness per raw bit is $r_{\mathrm{rand}}/r_{\mathrm{total}}$, the rest $1-r_{\mathrm{rand}}/r_{\mathrm{total}}$ bits are considered as classical noise, and, to be conservative, as to be known by Eve. Correspondingly, $p_{\mathrm{c}}=1-r_{\mathrm{rand}}/r_{\mathrm{total}}$, is the fraction of classical noise in the raw random data. The conditional min-entropy in this case can be written as:
\begin{equation}
\label{eqn:e_conmin3}
\begin{split}
&H_{\infty}(\textbf{X}|\textbf{Y})\\
&=-\mathrm{log}_2 \Big(p_{\mathrm{c}}+(1-p_{\mathrm{c}})\big(\sum_{\textbf{y}} p(\textbf{y}) \max\limits_{\textbf{x}}\{p(\textbf{x}|\textbf{y})\}\big)\Big)\, .
\end{split}
\end{equation}

%

\section{Experiment}

The above discussed schemes are experimentally realized. First, the experimental configuration is discussed. After this, the experimental subtleties are characterized and the entropy is estimated for the experimental data according to the three prior discussed cases.

\subsection{Experimental Implementation}

\begin{figure*}[htb]
  \includegraphics[width=\textwidth]{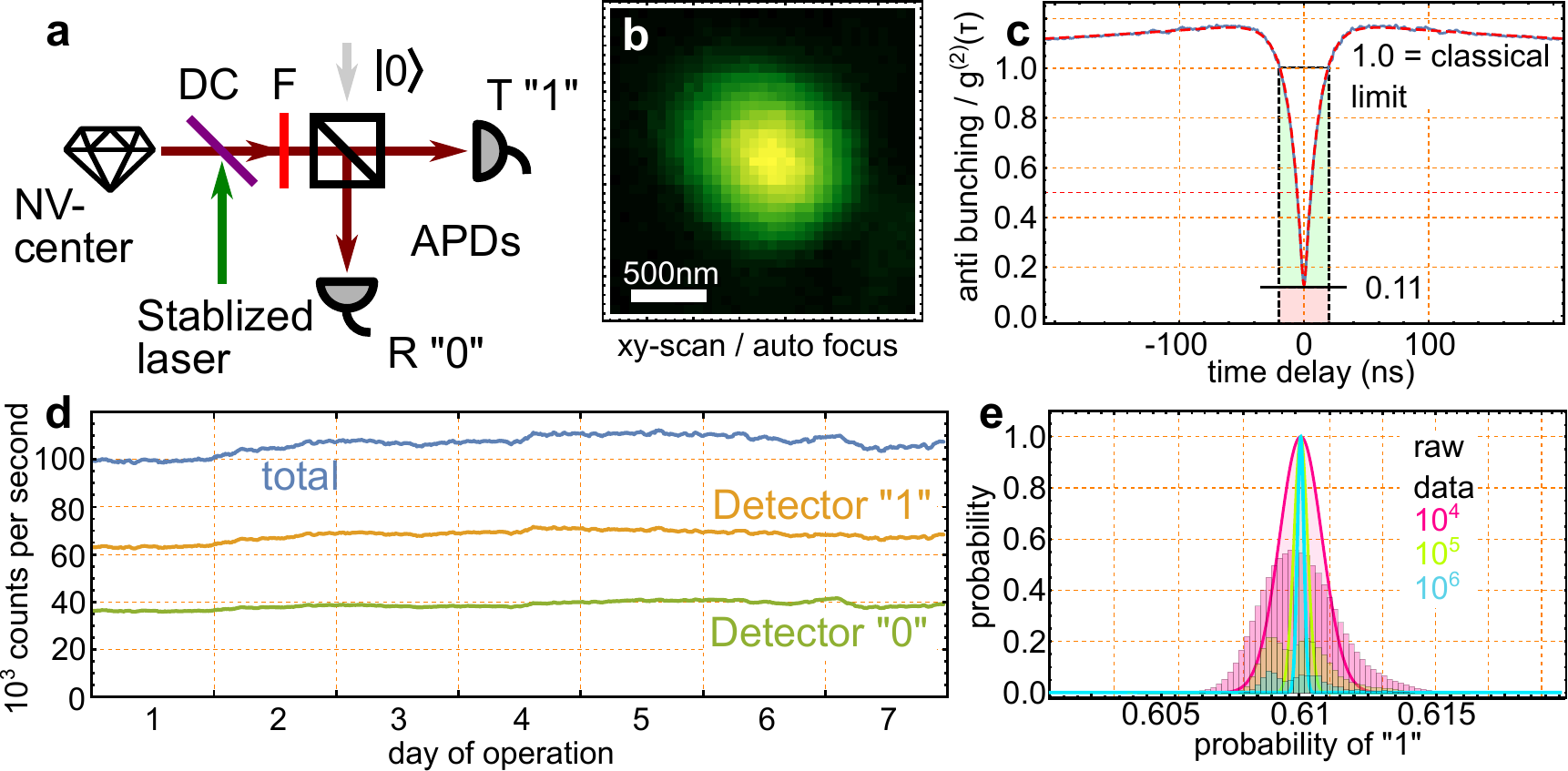}
  \caption{\textbf{Experimental Configuration.} \textbf{a}, Scheme of the experimental configuration. A confocal microscope is used to observe a single nitrogen-vacancy center. The detection is performed with two avalanche photo diodes (APDs). DC = Dichroic Mirror; F = Long-pass Filter. \textbf{b}, Fluorescence counts of a lateral scan over the sample. Peak intensity: 100~kcps (kilo counts per second). \textbf{c}, Measurement of anti-bunching and a theoretical fit (dashed line), the timing resolution here is 0.5~ns. \textbf{d}, a long time recording in the course of 7 days, the exact time is 608125 seconds. \textbf{e}, Presentation of the raw data and the experimental bias between 0 and 1.}
  \label{fig:fig01}
\end{figure*}

The single photon source used for randomness generation consists of the emission of a single nitrogen-vacancy center which is optically excited by a continuous wave laser. The resulting fluorescence is detected by confocal microscopy (simplified in Fig.~\ref{fig:fig01}a). The entire experiment is operated under ambient conditions, and spans less than 1~m$^2$ of an optical table.

The laser ($\lambda$=532~nm) which is used to excite the single emitter is operated in continuous wave mode, i.e.\ the utilized laser is always on. To avoid laser power fluctuations, the laser intensity is stabilized by a commercial PID-controller (Stanford Research, SIM960). For this, the laser power is detected shortly before the diamond single photon source. The laser power is regulated by an acousto-optical modulator which is located at the laser output. For mode-cleaning, the laser light is then guided by an optical single-mode fiber and introduced into the microscope.

After the laser beam is reflected off a dichroic mirror it is guided to a galvanometric mirror system which steers the beam to a 4$f$-scanning microscope. Focusing is realized by an 100$\times$ oil objective (Olympus Plan FL N, NA=1.35). In the confocal configuration, the emitted light is then captured by the same microscope objective, guided backwards, and transmits through the dichroic mirror towards the detection system. To suppress unwanted stray light, the detected light is then tightly focused ($f$=100~mm) onto a pinhole ($\varnothing$=50~$\upmu$m) and filtered by a 640~nm long pass filter. The detected light is then transferred by 2$f$-2$f$ imaging, through a symmetric non-polarizing beam splitter towards two single photon detectors (Count, Laser Components). This configuration reduces the avalanche photo diode (APD) cross talk significantly.

The sample is a mm-sized diamond which hosts nitrogen-vacancy-centers at natural abundance. For high excitation and collection efficiency of the NV-center a solid-immersion lens was fabricated around an earlier confocally localized center. Further details on its manufacture are presented elsewhere~\cite{jamali_rosi_2014}. The centers and the solid-immersion lenses are identified by confocal beam scanning. This was performed initially to locate the centers, but also in the course of the experiment the beam is repeatedly ($\Delta$t=8~min.) medially and laterally scanned across a certain area. Then, the NV-center is re-centered and the measurement is continued. This suppressed drift of the sample during the measurement time. One of the lateral images is presented in Fig.~\ref{fig:fig01}b.

All detection events are recorded on a commercial FPGA-based time-tagger (Swabian Instruments, Timetagger 20). The time-tagger is operated with a 100~ps time resolution and records all detector events. Since each produced click is recorded in a 128 Bit binary (64 bit which detector has clicked and 64 bit with the time in ps), we have recorded 832~GiB in the course of 7~days (to be specific: 608125~seconds). This data set is split to 179 files, which are analyzed below. This corresponds to an average count rate of approximately 91.7~kcps, which includes the refocusing periods, which display a reduced count-rate for the time of refocusing. Since the real-time count rate is about 100~kcps, then the time without refocusing is approximately 558000~s.

To prove that a single emitter has been observed and to show the single photon nature of the emitted photon stream, we analyze the anti-bunching of the photons in a Hanbury Brown and Twiss configuration (see also Fig.~\ref{fig:fig01}a). This is performed by correlating the recorded time-stamps of the two APDs in a start-multiple-stop fashion~\cite{oberreiter_light_2016}. The corresponding anti-bunching curve is shown in Fig.~\ref{fig:fig01}e. It shows an anti-bunching ``dip'' below the value of $g^{(2)}(0)$=0.5, which proves the single photon nature of the source. The timing resolution $\tau_{\mathrm{rs}}$ for the start-stop event is 500~ps. Furthermore, the curve shows some bunching behavior due to the NV-centers' typical meta-stable state above the low excitation limit.

The entire experiment was prepared to operate without any human interaction. During the experiment the entire setup was covered with black out material. The above mentioned refocusing procedure helped to ensure reliable operation. A measurement of the peak count rates is presented in Fig.~\ref{fig:fig01}d. Still, some fluctuations are observed in the course of the recording. These are caused mostly by thermal drift of the table. Most notably the position of the pinhole, and both APDs are influenced. The relative position of the excitation focus also plays an important role. We like to note that the average count rate (above) was only measured as 91~kcps and not approximately 100~kcps as shown here. This results from the fact that the refocusing procedure reduces the count rate periodically.

The single photon nature is online monitored by the recording of the auto-correlation function. This is implemented such that each photon click of one channel is correlated with all clicks in a certain time span of the other detector. This is a start-multiple-stop correlation, which does not go down as other recordings which only consider start-stop events~\cite{oberreiter_light_2016}.



\begin{figure*}[htb]
  \includegraphics[width=\textwidth]{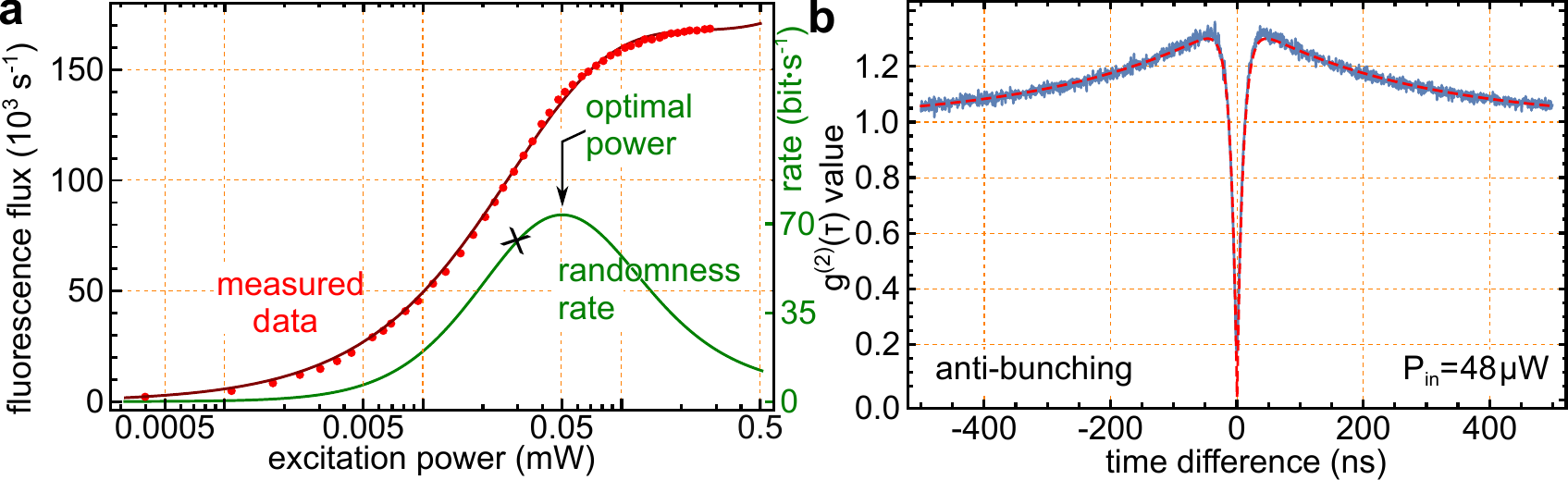}
  \caption{\textbf{Entropy and Randomness Estimation.} \textbf{a)}, Saturation curve of the utilized NV-center. To note the non-trivial behavior at higher laser powers, which indicate that the NV-center can not be considered as a simple three-level system. The optimal rate for the randomness generation speed is shown in green. This curve forms due to the fact that the anti-bunching curve gets narrower with an increasing laser power. This implies although more events are generated, the overall area below the curve is reduced. The cross in the green curve is the excitation power of the experimental data analyzed in supplementary material.  \textbf{b)}, the anti-bunching curve at the optimal point of the randomness generator. The bottom at $\tau$=0 amounts to $g^{(2)}(0)$=0.15.}
  \label{fig:fig02}
\end{figure*}

\subsection{Experiment Results}

In the course of 7 days we have acquired 832~GiB data. This corresponds to 55796707904~bits raw data. And the detected photons in the reflected arm are associated with the outcome 1, while the transmitted one are associated with the outcome 0. Then the number of zeros is 21753096536~bits, and the number of ones is 34043611368~bits. The integrated imbalance of the beam-splitter ratio and the detector efficiency amounts to probability $p(1)$=0.6101, $p(0)$=0.3899. This bias is indicated in Fig.~\ref{fig:fig01}e. Still, this bias is not necessarily a problem, since we assume the bits as still to be independent from each other. This is justified since the average waiting time between two photon detection events is considerably far away from the dead-time of the utilized single photon detectors. Furthermore, the given imbalance does only lead to a few percent reduction of the final entropy rate.

Still, this bias will largely decrease the usability of the random bits. We need to post-process the raw randomness bits and make them well balanced. When unbiasing the raw bits by two universal hashing, the conditional min-entropy is calculated by Eqn.~(\ref{eqn:condmin}). $H_{\infty}(X|Y)$ gives us a conservative bound, which is how many random bits can be extracted per raw bit.

\begin{figure*}[htb]
  \includegraphics[width=\textwidth]{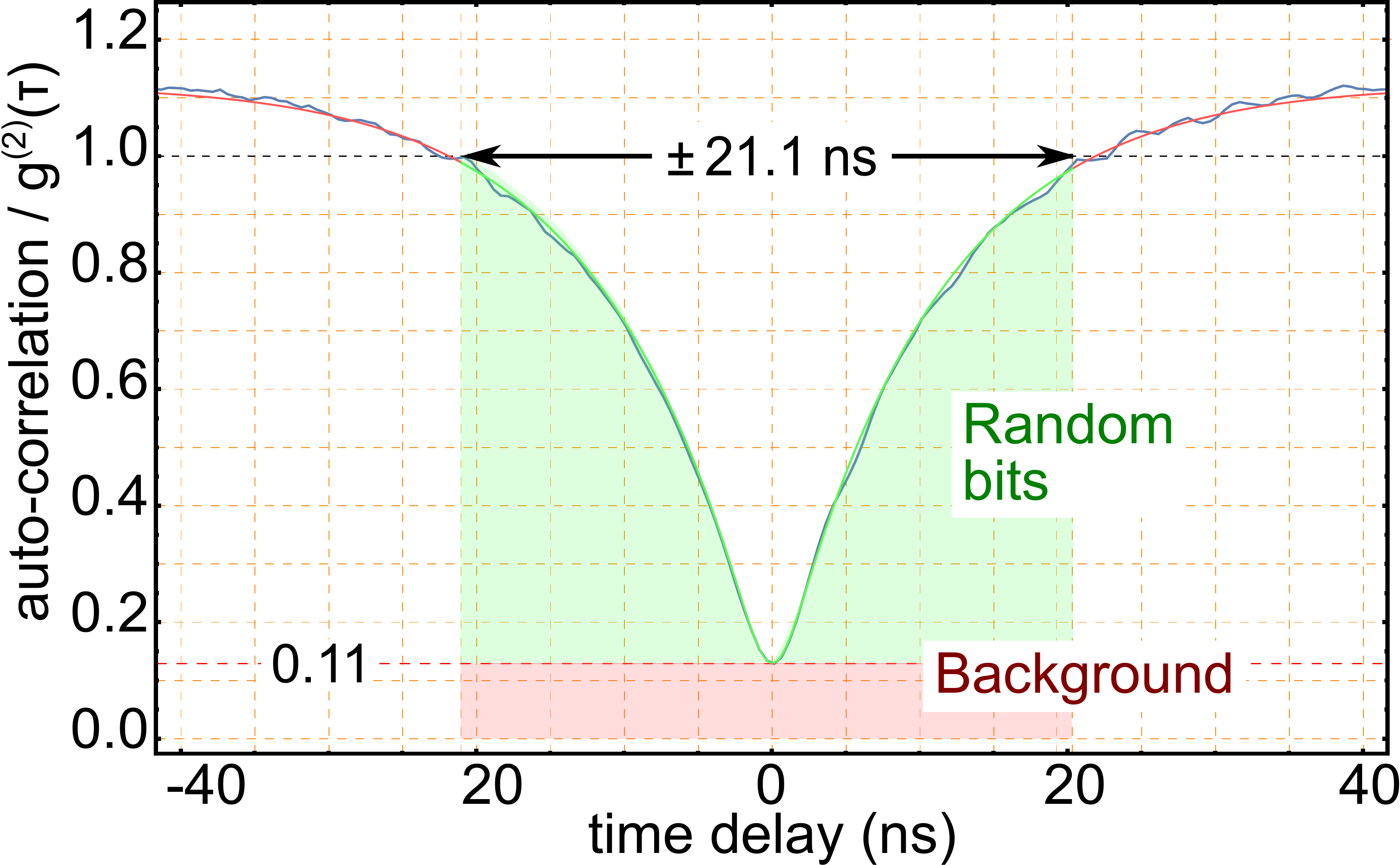}
  \caption{\textbf{Anti-bunching as a Measure for Quantumness.} The anti-correlation of photons is only observed in a small time interval. For the third randomness extraction model, the area of the generated bits between the classical bound of $g^{(2)}(\tau)\leq$1.0 and above the background level are considered. This reduces the amount of raw input bits for the generator dramatically.}
  \label{fig:fig03}
\end{figure*}

In our case, when considering the 11.5$\sigma$ error bound, $H_{\infty}(X|Y)$ is 0.5559~bits, which means for per raw bit, 0.5559~bits secured random number can be extracted. With this value, by two universal hashing, we can get approximate $3.10\times10^{10}$ unbiased random bits from the raw bits. The output speed of these unbiased random bits is $5.10\times10^{4}$ bits per second, when the refocusing periods where the count-rate is lowered are still counting as randomness generation time.

By limiting the generated raw random data to single photon events in the second model, we can guarantee the independence from uncorrelated background in the random data. Using Eqn.~(\ref{eqn:e_conmin}), with 11.5$\sigma$ error bound, the extractable quantum randomness from per raw bit amounts to 0.5168 bits. For the whole raw bits, the extractable quantum random bits are $2.88\times10^{10}$ bits. When including the refocusing periods, the output speed of the quantum random number generator amounts to $4.74\times10^4$ bits per second.

Note that this quantum random number generation speed is smaller than the unbiased random bits generation speed in the first model. The difference between these two model indicates that in the first model, some classical noise background might have been considered as random events.

Next, we calculate the extractable quantum random bits from the perspective of the third model, which gives us some extended independence of an eavesdropper.

The fluorescence counts and the shape of anti-bunching curves are affected by the excitation power. Subsequently, $R_{\mathrm{rand}}$ depends on the excitation power to the single quantum emitter. As shown in Fig.~\ref{fig:fig02}a, the green curve is the quantum randomness output rate; it depends on different excitation powers. The curve has an optimal excitation power as expected. This stems from the fact that with an increase of the excitation power, the count rate of different detectors increases, while the shape of the anti-bunching curves becomes narrower, thus the green part in Fig.~\ref{fig:fig03} would become smaller. At the given excitation intensity of 26$\mu$W, the green part covers a range of $t=$10.55~ns. When the excitation power is changed, the start-stop event count rate will first increase and later decrease. Subsequently, the quantum random bits output rate has an optimal operation point. For a simple three level system this rate matches with the saturation point of the utilized single photon emitter.

According to Eqn.~(\ref{rand}), we can calculate the quantum randomness output rate of our system before we post-process the raw random bits. The excitation power in the experiment is $26$~$\mu$W. The integrated ratio of the beam splitter and the detection efficiency amounts as above to about 0.6101.

Following Eqn.~(\ref{eqn:e_conmin3}), with a very strict 11.5$\sigma$ error bound, we compute the extractable quantum randomness per raw bit as $3.746\times10^{-4}$. With this value, after the randomness-extraction hashing, about $2.09\times10^7$ bits unbiased quantum random number can be extracted, and the quantum random number generation speed amounts to approximately $34.37$ bits per second.

\section{Conclusion and Outlook}
In conclusion, we have theoretically described and experimentally implemented a random bit generator based on single photons which are impinging on a beam-splitter. The utilized single photon source is based on a single defect center in diamond. The generator is operated continuously over the course of one week and all detector events are recorded as time-tags, such that they could be conveniently post-processed.

The detection of raw random bits, which are associated with the output ports of the beam-splitter to ones and zeros result in a raw-bit stream. This has a number of subtleties. The single photon detection process is prone to technical effects such as the beam-splitter ratio, electrical dead-times and jitter. As with any other randomness generator, the raw bits can subsequently not been used without post-processing. Therefore an entropy analysis for the raw random bit stream was presented. In a further analysis, this model can be extended to estimate the amount of unwanted, and potentially untrusted background events. This estimation is based on the parameters of the recorded photon correlation function.

In a third method, only tuple detection events on changing bits are considered as raw bits. The limitation is further reduced to auto-correlation values below unity, and excluded the uncorrelated background. This selection, a subset of detection events, certifies the quantum nature of the source. Despite this quantum nature, the ``decision'' the experimental outcome is based on the fair-sampling assumption of the beam-splitter. Therefore, the fundamental randomness process is not tied to the quantum nature of the source. In this sense also any other input light source might sample the vacuum fluctuations at the empty beam-splitter port, which will be the relevant entropy source of such beam-splitter based generators, while the quantum input state only certifies the ``quantumness'' of the utilized light source.

By an estimation of the underlying entropy of this conservative model, which is bound to the knowledge of an external adversary, we can estimate the quantum randomness per raw bit is $3.746\times10^{-4}$ bits. This fraction is used as input parameter for a randomness extraction. Then, the random bits are extracted. The final quantum random number generation speed is then about 34.37 bits per second.

\section*{Acknowledgements}
We acknowledge useful comments from Prof.\ R.\ Renner, ETH Zurich. We further acknowledge the funding from the MPG, the SFB project CO.CO.MAT/TR21, ERC (Complexplas), the BMBF, the Eisele Foundation, the project Q.COM, and SMel. Also the network IQ$^{ST}$ is acknowledged for funding.

\section*{Author contributions statement}
I.G. and J.N.G. conceived the experiment(s),  I.G. and X.C. conducted the experiments, X.C. analysed the results.  All authors reviewed the manuscript. 

\section*{Additional information}
The authors declare no competing interests.

\newpage

\setcounter{section}{0}
\noindent
{\raggedright\sffamily\bfseries\fontsize{18}{22}\selectfont Supplementary material to:\\Single Photon Randomness based on a Defect Center in Diamond\par}%
\vskip10pt
{\raggedright\sffamily\fontsize{10}{14}\selectfont  Xing Chen, Johannes Greiner, J\"org Wrachtrup and Ilja Gerhardt\par}
\vskip18pt%

\noindent
This supplementary material provides details about some of the concepts and equations used in the main text. In the first section, the basic equations of the three level system and the photon statistics are introduced. In the second section, the data post-processing methods are described, including the derivation of the conditional min-entropy, the error bound of the conditional min-entropy, and two-universal hashing. In the third section, our experimental data is analysed.

\section{Three Level System}
\begin{figure*}[h]
\begin{center}
\includegraphics[width=4cm]{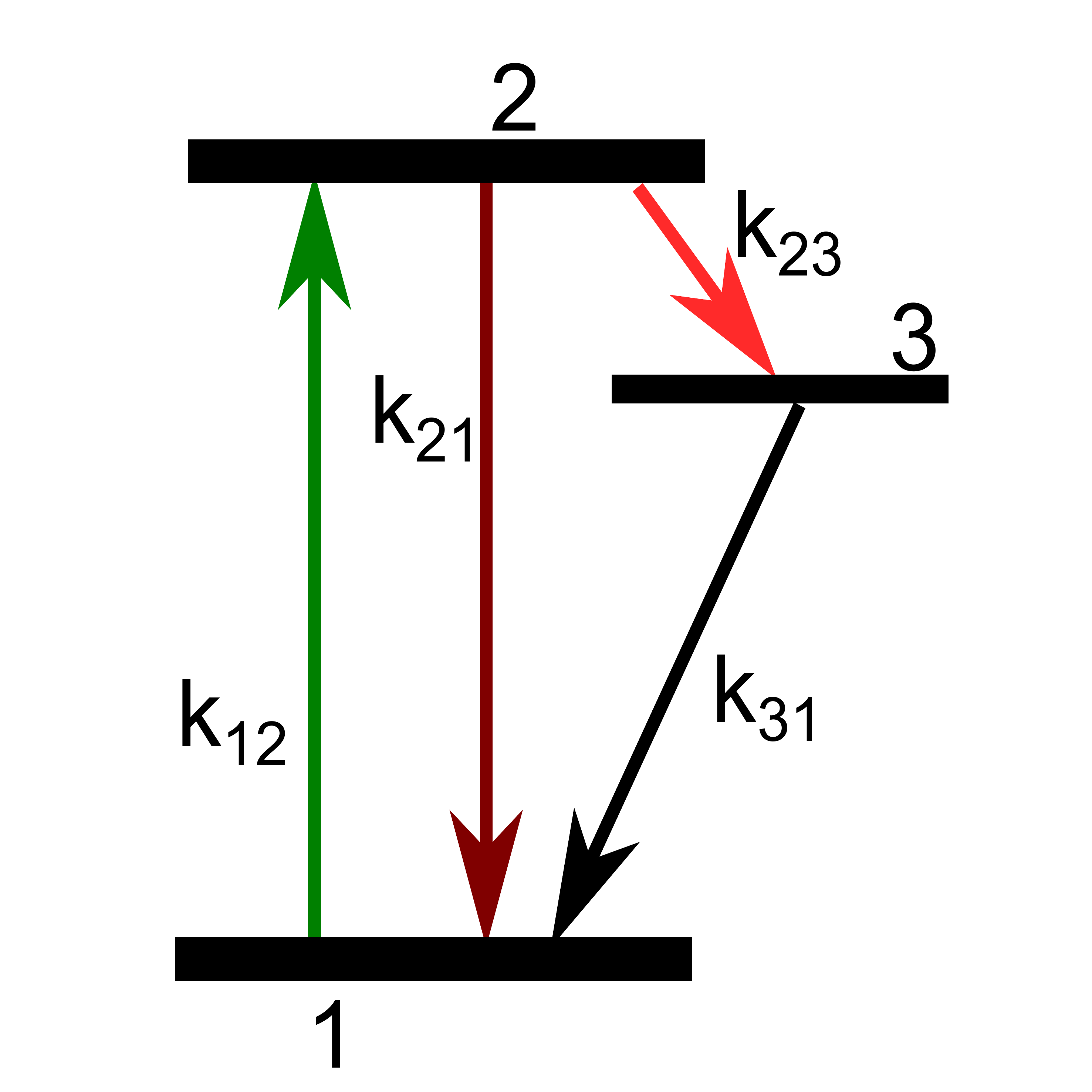}
\end{center}
\caption{Three level system. The ground state is 1, the excited state is 2, and 3 is a meta-stable state.}
\label{fig:fig01_threelevel}
\end{figure*}

The single photon source in our experimental setup is based on a single nitrogen-vacancy (NV) centre. This can be described as a three-level system. Please refer to Fig.~\ref{fig:fig01_threelevel} for a naming of the levels. The rate equations for this three level system are
\begin{equation}\nonumber
\begin{split}
&\dot{\rho_1}=-k_{12}\rho_1+k_{21}\rho_2+k_{31}\rho_3~,\\
&\dot{\rho_2}=k_{12}\rho_1-(k_{21}+k_{23})\rho_2~,\\
&\dot{\rho_3}=k_{23}\rho_2-k_{31}\rho_3~.
\end{split}
\end{equation}
where $\rho_{i}$ denotes the population of each state, and $\dot{\rho_i}$ indicates the time derivative of $\rho_i$, $k_{12}$ is the pumping rate, the other $k_{ij}$ are the decay rates. Without excitation power, the NV centre is in its ground state, so the initial conditions are $\rho_1=1, \rho_2=\rho_3=0$. By solving this differential equation system, we get $\rho_1(\tau),\rho_2(\tau),\rho_3(\tau)$. The auto-correlation function $g^{(2)}$ is defined as
\begin{equation}
g^{(2)}(\tau)=\frac{\rho_2(\tau)}{\rho_2(\infty)}~.
\label{eqn:g2}
\end{equation}

The average fluorescence rate under continuous wave(CW) excitation is
\begin{equation}\nonumber
F=\rho_2(\infty)k_{21}~.
\end{equation}

The different parameters used to fit the experimental saturation curve, as shown in Fig.2a in the main text, are
\begin{equation}\nonumber
\begin{split}
&k_{12}=0.77601~\mathrm{mW}^{-1} P_{\mathrm{in}}~\mathrm{ns}^{-1}~,\\
&k_{21}=\frac{1}{16.08}~\mathrm{ns}^{-1}~,\\
&k_{23}=\left(\frac{1}{30.97}~\mathrm{mW}^{-1}P_{\mathrm{in}}+\frac{1}{1000}\right)~\mathrm{ns}^{-1}~,\\
&k_{31}=\left(\frac{1}{78.37}\mathrm{mW}^{-1}P_{\mathrm{in}}+\frac{1}{471.7}\right)~\mathrm{ns}^{-1}~.
\end{split}
\end{equation}
where $P_{\mathrm{in}}$ is the excitation power, unit is mW, and refers to the power at the entrance of the microscope objective. Theoretically, $k_{12}$ equals $P_{\mathrm{in}}$, but realistically, the excitation power cannot be used without loss, and that is why there is an efficiency parameter between $k_{12}$ and $P_{\mathrm{in}}$.

The relationship between the background noise events and the excitation power $P_{\mathrm{in}}$ is considered as linear, and their relationship is found to be
\begin{equation}\nonumber
BG_{\mathrm{rate}}=5.2\times10^{6}~\mathrm{mW}^{-1}P_{\mathrm{in}}~\mathrm{s^{-1}}~.
\end{equation}

With background noise, the total photon events rate $F_{\mathrm{bg}}$ is
\begin{equation}\nonumber
\begin{split}
F_{\mathrm{bg}}&=F+BG_{\mathrm{rate}}\\
&=\rho_2(\infty)k_{21}+5.2\times10^{6}~\mathrm{mW}^{-1}P_{\mathrm{in}}~\mathrm{s^{-1}}~.
\end{split}
\end{equation}

The overall efficiency amounts to approximately 0.779\% in our setup, this value denotes the ratio of detected photon events and the total photons events $F_{\mathrm{bg}}$. This overall efficiency is affected by multiple parameters, including the transmission efficiencies of the lens, the reflection coefficients of the mirrors, and the detection efficiencies of the detectors.

Our data was recorded at $P_{\mathrm{in}}=26\mathrm{m W}$, and according to the above equations, $F_{\mathrm{bg}}=1.28\times10^7\mathrm{s}^{-1}$. Since the overall efficiency here is 0.779\%, the detection rate is about
\begin{equation}\nonumber
1.28\times10^7\mathrm{s}^{-1}\times 0.779\%=99.7~\mathrm{kcps}~.
\end{equation}

From~\ref{eqn:g2} and~\cite{Brouri:00_sup}, the equation for $g^{(2)}_{\mathrm{fit}}(\tau)$ is derived as
\begin{equation}
g^{(2)}_{\mathrm{fit}}(\tau)=1-1.06678 e^{-8.66533\times10^7 \left| \tau \right| \mathrm{s}^{-1}}+0.178378 e^{-2.00826\times10^6 \left| \tau \right| \mathrm{s}^{-1}}~.
\label{eqn:g2fit}
\end{equation}

$g^{(2)}_{\mathrm{fit}}(\tau)$ is the formula of anti-bunching curve that fits our experimental data.
\section{Data Post-processing}
When we post-process the experimental data, there are four tuples 00, 01, 10, 11, and their corresponding probabilities are p(00), p(01), p(10), p(11).

\subsection{Tuples}

To calculate the conditional probabilities on a given set of experimental data, we count these tuple events in an overlapping fashion. This means, for example, for a number string like $010$, when the tuples are overlapping, this string contains tuples $01$ and $10$; in the case of non-overlapping counting, this string only contains one tuple: $01$.\\

The number of the overlapped tuples is calculated in the following way. Suppose we have a random string with $N$ bits, when this random string is perfectly balanced, we should have $N(0)=N(1)$, this means that $N$ at least should be an even number. Also, in a perfectly balanced random string, the number of tuples should satisfy two conditions: $N(00)=N(01)=N(10)=N(11)$ and $N(00)+N(01)+N(10)+N(11)=N$. For a random string with $N$ bits, in order to satisfy $N(00)=N(01)=N(10)=N(11)$, $N$ must be a number which can be divided by 4. \\

For any perfectly balanced random string, $N(00)+N(01)+N(10)+N(11)=N$ cannot be satisfied by a finite $N$, because for any finite $N$, the total number of tuples always adds up to $N-1$ instead of $N$. In order to make $N(00)+N(01)+N(10)+N(11)=N$, when we calculate the number of overlapped tuples, we put the last bit $n_l$ and the first bit $n_f$ together to get an additional tuple $n_l n_f$, and by this way the condition $N(00)+N(01)+N(10)+N(11)=N$ will be fulfilled. For example, for a random string like $1110101000001110$, the length of this string is $N=16$, while $N(00)=4, N(01)=3, N(10)=4, N(11)=4$, the total number of tuples is 15. When we put the last bit $0$ and the first bit $1$ together as a new tuple $01$, it will make $N(01)=4$, and they sum up to the desired number 16. We consider this as the optimal way to treat measured data and to calculate the frequencies of the single and double events.\\

\subsection{The deduction of the conditional min-entropy}
In this subsection, the conditional min-entropy mentioned in the main text is calculated in detail. The conditional min-entropy is defined as
\begin{equation}\nonumber
H_{\infty}(X|Y)=-\mathrm{log}_2 \big(\sum_y p(y) \max\limits_{x}\{p(x|y)\}\big)~.
\end{equation}
where $x$ and $y$ are two subsequent random bits in a given random string.

In our case, $\{X,Y\}\in \{0,1\}$, subsequently the conditional min-entropy is
\begin{equation}
\begin{split}
H_{\infty}(X|Y)&=-\mathrm{log}_2 \big(\sum_y p(y) \max\limits_{x}\{p(x|y)\}\big)\\
&=-\mathrm{log}_2\big(p(0)\max\{p(0|0),p(1|0)\}+p(1)\max\{p(0|1),p(1|1)\}\big)~.
\label{eqn:Hinf}
\end{split}
\end{equation}

For the convenience of description, the detector in the transmitted arm is named as detector A, and the detector in the reflected arm is named as detector B. Next, we deduce all the conditional probabilities from the experimental parameters. We note that this does not describe the true probabilities, but reflects the frequencies of the occurring singles and tuples.\\

Take $p(A|A)$ as an example, $p(A|A)$ means the probability of a subsequent photon to be detected in detector A when detector A has already detected a previous photon event. Let $\eta_{\mathrm{A}}$ be the detection efficiency of detector A, $\tau^{\mathrm{A}}_{\mathrm{dead}}$ be the dead-time of detector A, and $T$ be the transmission coefficient of the beam-splitter. When detector A clicked, it is in its dead-time. $\int_0^{\tau^{\mathrm{A}}_{\mathrm{dead}}}g^{(2)}_{\mathrm{fit}}(\tau)d\tau$ means the probability that the next incident photon is in the dead-time of detector A, then the probability of this incident photon outside its dead-time is $1-\int_0^{\tau^{\mathrm{A}}_{\mathrm{dead}}}g^{(2)}_{\mathrm{fit}}(\tau)d\tau$. When the incident photon is outside the dead-time of detector A, it has probability $T$ to be transmitted to detector A, and detector A has probability $\eta_{\mathrm{A}}$ to detect this photon, so $p(A|A)$ could be written as
\begin{equation}
p(A|A)=\big(1-\int_0^{\tau^{\mathrm{A}}_{\mathrm{dead}}}g^{(2)}_{\mathrm{fit}}(\tau)d\tau\big)\eta_{\mathrm{A}}T~.
\label{eqn:p(0|0)}
\end{equation}

Similarly, the parametric equation for $p(B|A)$ is
\begin{equation}
p(B|A)=\eta_{\mathrm{B}}R\big(1-\underbrace{\eta_{\mathrm{B}}R\int_0^{\frac{\tau_{\mathrm{dead}}^{\mathrm{B}}}{2}}g^{(2)}_{\mathrm{fit}}(\tau)d\tau\int_0^{\frac{\tau_{\mathrm{dead}}^{\mathrm{B}}}{2}}g^{(2)}_{\mathrm{fit}}(\tau)d\tau}_{\text{probability that detector B is in its dead-time}}\big)~.
\label{eqn:p(1|0)}
\end{equation}

where $\eta_{\mathrm{B}}$ is the detection efficiency of detector B, $\tau^{\mathrm{B}}_{\mathrm{dead}}$ is the dead-time of detector B, and $R$ is the reflection coefficient. This gives us the equation of $p(B|A)$, which means when detector A detects a photon event, the probability of detector B detecting a subsequent photon event. \\

The formula in the underbrace means the probability of detector B is not in its dead-time when a photon shoots into the beam-splitter, this probability is an estimation, which is based on the assumption $\tau^{\mathrm{A}}_{\mathrm{dead}}\approx\tau^{\mathrm{B}}_{\mathrm{dead}}$. Before detector A clicks, the previous photon event may be on detector A or B. If it is on detector A, it will not affect the conditional probability $p(B|A)$, since when detector A clicks two times, detector B is ready to detect a photon event; if the previous photon event is on detector B, then after detector A's click, detector B still has a probability to be in its dead-time when the next photon comes into the beam-splitter. Inside the brace, the half of the dead-time of detector B is a simplified version of the above probability analysis, where $\eta_{\mathrm{B}}R\int_0^{\frac{\tau_{\mathrm{dead}}^{\mathrm{B}}}{2}}g^{(2)}_{\mathrm{fit}}(\tau)d\tau$ gives us the probability that the previous photon event fires on detector B within the half of the dead-time of detector B, and $\int_0^{\frac{\tau_{\mathrm{dead}}^{\mathrm{B}}}{2}}g^{(2)}_{\mathrm{fit}}(\tau)d\tau$ is the probability that the next incident photon is inside the half of the dead-time of detector B.\\

$p(B|B)$ and $p(A|B)$ can be derived analogously.\\

The equation of $p(A)$ is $p(A)=r_{\mathrm{A}}/r_{\mathrm{total}}$, $r_{\mathrm{A}}$ is the click rate of detector A. $r_{\mathrm{A}}$ is defined as following
\begin{equation}
\begin{split}
r_{\mathrm{A}}=&\eta_{\mathrm{A}}T I_{\mathrm{in}}-\underbrace{\frac{(\eta_{\mathrm{A}}T I_{\mathrm{in}})}{2}\times\frac{(\eta_{\mathrm{A}}T I_{\mathrm{in}})}{2}\int_0^{\tau_{\mathrm{dead}}^{\mathrm{A}}}g^{(2)}_{\mathrm{fit}}(\tau)d\tau}_{\text{rate of two clicks within the dead-time of detector A}}\\
&=\eta_{\mathrm{A}}T I_{\mathrm{in}}-\frac{(\eta_{\mathrm{A}}T I_{\mathrm{in}})^2\int_0^{\tau_{\mathrm{dead}}^{\mathrm{A}}}g^{(2)}_{\mathrm{fit}}(\tau)d\tau}{4}~.
\end{split}
\label{eqn:rt}
\end{equation}

where $I_{\mathrm{in}}$ is the rate of incident photon, and $\eta_{\mathrm{A}}T I_{\mathrm{in}}$ means the click rate when detector A would have no dead-time. The latter part of the equation is the probability of two subsequent events in detector A to have a time distance which is smaller than the dead-time of detector A. For detector  B, a similar equation of $r_{\mathrm{B}}$ is derived, we get
\begin{equation}
\begin{split}
   p(A)&=\frac{r_{\mathrm{A}}}{r_{\mathrm{A}}+r_{\mathrm{B}}}\\
   &=\frac{\eta_{\mathrm{A}}T I_{\mathrm{in}}-\frac{(\eta_{\mathrm{A}}T I_{\mathrm{in}})^2\int_0^{\tau_{\mathrm{dead}}^{\mathrm{A}}}g^{(2)}_{\mathrm{fit}}(\tau)d\tau}{4}}{\eta_{\mathrm{A}}T I_{\mathrm{in}}-\frac{(\eta_{\mathrm{A}}T I_{\mathrm{in}})^2\int_0^{\tau_{\mathrm{dead}}^{\mathrm{A}}}g^{(2)}_{\mathrm{fit}}(\tau)d\tau}{4}+\eta_{\mathrm{B}}R I_{\mathrm{in}}-\frac{(\eta_{\mathrm{B}}R I_{\mathrm{in}})^2\int_0^{\tau_{\mathrm{dead}}^{\mathrm{B}}}g^{(2)}_{\mathrm{fit}}(\tau)d\tau}{4}}\\
   &=\frac{\eta_{\mathrm{A}}T -\frac{(\eta_{\mathrm{A}}T)^2 I_{\mathrm{in}}\int_0^{\tau_{\mathrm{dead}}^{\mathrm{A}}}g^{(2)}_{\mathrm{fit}}(\tau)d\tau}{4}}{\eta_{\mathrm{A}}T-\frac{(\eta_{\mathrm{A}}T)^2 I_{\mathrm{in}}\int_0^{\tau_{\mathrm{dead}}^{\mathrm{A}}}g^{(2)}_{\mathrm{fit}}(\tau)d\tau}{4}+\eta_{\mathrm{B}}R-\frac{(\eta_{\mathrm{B}}R)^2 I_{\mathrm{in}}\int_0^{\tau_{\mathrm{dead}}^{\mathrm{B}}}g^{(2)}_{\mathrm{fit}}(\tau)d\tau}{4}}~.
\label{eqn:p(0)}
 \end{split}
\end{equation}

With all the above equations, the parametric expression of $H_{\infty}(X|Y)$ could be deduced.

\subsection{Error bound of the entropy}
\label{subsec:error_bound}
In this subsection, the error bound of the conditional min-entropy is given. Let us mention some properties of the error bound. Since $p(A)+p(B)=1$ is always fulfilled, we have $\Delta_{p(A)}=-\Delta_{p(B)}$. Also, for conditional probabilities $p(A|A),p(B|A),p(A|B),p(B|B)$, $p(A|A)+p(B|A)=1$ and $p(A|B)+p(B|B)=1$, this means $\Delta_{p(A|A)}=-\Delta_{p(B|A)}$ and $\Delta_{p(A|B)}=-\Delta_{p(B|B)}$. Since $p(AB)=p(BA)$($p(BA)=p(AB)$ is satisfied under the condition that the experimental devices does not change over time), and $\Delta_{p(A)}=-\Delta_{p(B)}$, it is easy to derive the relationship  $\Delta_{p(A|A)}=-\Delta_{p(B|A)}=\Delta_{p(A|B)}=-\Delta_{p(B|B)}$.\\

The conditional min-entropy in our case is defined in~\ref{eqn:Hinf}. Since $p(BA)=p(AB)$,
\begin{equation}
H_{\infty}(X|Y)=-\mathrm{log}_2\big(\max\{p(A)-p(AB),p(AB)\}+\max\{p(AB),1-p(A)-p(AB)\}\big)~.
\label{eqn:H_inf1}
\end{equation}

There are four different conditions for $H_{\infty}(X|Y)$
\begin{equation}
H_{\infty}(X|Y)=\begin{cases}
      -\mathrm{log}_2(p(A)) & p(A)-p(AB)\geq p(AB)\quad and\quad p(AB)\geq 1-p(A)-p(AB)~,\\
      -\mathrm{log}_2(p(B)) & p(A)-p(AB)\leq p(AB)\quad and\quad p(AB)\leq 1-p(A)-p(AB)~,\\
      -\mathrm{log}_2(2p(AB)) & p(A)-p(AB)\leq p(AB)\quad and\quad p(AB)\geq 1-p(A)-p(AB)~,\\
      -\mathrm{log}_2(1-2p(AB)) & p(A)-p(AB)\geq p(AB)\quad and\quad p(AB)\leq 1-p(A)-p(AB)~.
   \end{cases}
   \label{eqn:H_inf conditions}
\end{equation}

No matter which condition $H_{\infty}(X|Y)$ is in, there is only one variable in it. Then a more conservative conditional min-entropy could be written as
\begin{equation}
H_{\infty}(X|Y)=-\mathrm{log}_2(f(p)+\Delta_{f(p)})
\label{eqn:con_H_inf_1}
\end{equation}

where $f(p)=\mathrm{max}\{p(A),p(B),2p(AB),1-2p(AB)\}$.\\

The equation of $p(A)$ is Eqn.~\ref{eqn:p(0)}, $p(A)$ is affected by the transmission coefficient $T$, the rate of incident photon $I_{\mathrm{in}}$, the detection efficiency $\eta_{\mathrm{A}}$, $\eta_{\mathrm{B}}$, and the dead-time $\tau_{\mathrm{dead}}^{A}$, $\tau_{\mathrm{dead}}^{B}$ of the two detectors. The error bound of each parameter is, $\delta_T$, $\delta_{\eta_{\mathrm{A}}}$,$\delta_{\eta_{\mathrm{B}}}$, $\delta_{\tau_{\mathrm{dead}}^{A}}$,$\delta_{\tau_{\mathrm{dead}}^{B}}$, and $\delta_{I_{\mathrm{in}}}$. According to the error propagation, the error bound of $p(A)$ is
\begin{equation}
\begin{split}
\Delta_{p(A)}=\bigg((\frac{\partial p(A)}{\partial T}\delta_T)^2+(\frac{\partial p(A)}{\partial \eta_{\mathrm{A}}}\delta_{\eta_{\mathrm{A}}})^2+(\frac{\partial p(A)}{\partial \eta_{\mathrm{B}}}\delta_{\eta_{\mathrm{B}}})^2+\\
(\frac{\partial p(A)}{\partial \tau_{\mathrm{dead}}^{A}}\delta_{\tau_{\mathrm{dead}}^{A}})^2+(\frac{\partial p(A)}{\partial \tau_{\mathrm{dead}}^{B}}\delta_{\tau_{\mathrm{dead}}^{B}})^2+(\frac{\partial p(A)}{\partial I_{\mathrm{in}}}\delta_{I_{\mathrm{in}}})^2\bigg)^{\frac12}~.
\end{split}
\label{eqn:delta_p(0)}
\end{equation}

The equation of $p(AB)$ is $p(AB)=p(A)p(B|A)$. From~\ref{eqn:p(1|0)} and~\ref{eqn:p(0)}, we know that multiple parameters affect $p(AB)$, including $\tau_{\mathrm{dead}}^{A}$, $\tau_{\mathrm{dead}}^{B}$,  $\eta_{\mathrm{A}}$, $\eta_{\mathrm{B}}$, $T$, and $I_{\mathrm{in}}$, similarly, $\Delta_{p(AB)}$ is
\begin{equation}
\begin{split}
\Delta_{p(AB)}=\bigg((\frac{\partial p(AB)}{\partial T}\delta_T)^2+(\frac{\partial p(AB)}{\partial \eta_{\mathrm{A}}}\delta_{\eta_{\mathrm{A}}})^2+(\frac{\partial p(AB)}{\partial \eta_{\mathrm{B}}}\delta_{\eta_{\mathrm{B}}})^2+\\
(\frac{\partial p(AB)}{\partial \tau_{\mathrm{dead}}^{A}}\delta_{\tau_{\mathrm{dead}}^{A}})^2+(\frac{\partial p(AB)}{\partial \tau_{\mathrm{dead}}^{B}}\delta_{\tau_{\mathrm{dead}}^{B}})^2+(\frac{\partial p(AB)}{\partial I_{\mathrm{in}}}\delta_{I_{\mathrm{in}}})^2\bigg)^{\frac12}~.
\end{split}
\label{eqn:delta_p(1|0)}
\end{equation}

Then from Eqn.~\ref{eqn:con_H_inf_1},~\ref{eqn:delta_p(0)},~\ref{eqn:delta_p(1|0)}, the conservative $H_{\infty}(X|Y)$ could be calculated.\\

For the second model, there is one more parameter $p_e$, which represents the probability of detecting an uncorrelated background noise event. The conditional min-entropy for the second model is
\begin{equation}
\begin{split}
H_{\infty}(X|Y)&=-\mathrm{log}_2 \Big(p_e+(1-p_e)\big(\sum_y p(y) \max\limits_{x}\{p(x|y)\}\big)\Big)\\
&=-\mathrm{log}_2 \Big(p_e+(1-p_e)f(p)\Big)~.
\end{split}
\label{eqn:H_inf 2}
\end{equation}

where $f(p)=\mathrm{max}\{p(A),p(B),2p(AB),1-2p(AB)\}$. A more conservative $H_{\infty}(X|Y)$ for this model is
\begin{equation}
H_{\infty}(X|Y)=-\mathrm{log}_2 \big(p_{eq}+\Delta_{p_{eq}}\big)
 \label{eqn:con_H_inf_2}
\end{equation}

where $p_{eq}=p_e+(1-p_e)f(p)$, then $\Delta_{p_{eq}}=\sqrt{(\frac{\partial p_{eq}}{\partial p_e}\Delta_{p_e})^2+(\frac{\partial p_{eq}}{\partial f(p)}\Delta_{f(p)})^2}$, where $\Delta_{f(p)}$ is derived from~\ref{eqn:delta_p(0)} and~\ref{eqn:delta_p(1|0)}, and $\Delta_{p_e}$ is shown in~\ref{eqn:delta_p_e}.

\subsection{Fraction of single photon events}
In the second model, regarding the fraction of the single photon events, we assume that the background noise is uncorrelated, and the single photon events are randomly mixed with the uncorrelated background noise. In this scenario, let the fraction of single photon events be $s$, then the fraction of the uncorrelated background noise is $1-s$. The fraction of single photon events is~\cite{Brouri:00_sup}
\begin{equation}
s=\sqrt{1-g^{(2)}_{\mathrm{fit}}(0)}~.
\label{eqn:single_fraction_s1}
\end{equation}

Of course, the deviation of the $g^{(2)}_{\mathrm{fit}}(0)$ from 0 does not mean background noise all the time, it may also be caused by multiple emitters~\cite{loudon2000quantum_sup}. The multiple equally bright emitters will change $g^{(2)}_{\mathrm{fit}}(0)$ to $1-\frac1n$, where $n$ is the number of emitters. For example, when $g^{(2)}_{\mathrm{fit}}(0)=0.5$, it could be caused by background noise or two emitters. When it is caused by two emitters, all the events are from single photon sources; when it is caused by background noise, from~\ref{eqn:single_fraction_s1}, we know that only 70.7\% events are single photon events. From a conservative consideration, as long as $g^{(2)}_{\mathrm{fit}}(0)$ deviates from 0, we treat it as the result of background noise, instead of some extra emitters.\\

Note that the uncertainty of $g^{(2)}_{\mathrm{fit}}(0)$ mentioned above will again affect the fraction of background noise, thus affect $p_e$. Since $p_e=1-s=1-\sqrt{1-g^{(2)}_{\mathrm{fit}}(0)}$, according to the propagation of uncertainty, the uncertainty of $p_e$ is
\begin{equation}
\Delta_{p_e}=\frac{1}{2\sqrt{1-g^{(2)}_{\mathrm{fit}}(0)}}\Delta_{g^{(2)}_{\mathrm{fit}}(0)}~.
\label{eqn:delta_p_e}
\end{equation}
where $\Delta_{g^{(2)}_{\mathrm{fit}}(0)}$ is derived in the following(see Eqn.~\ref{eqn:g2_fluctuation}).

\subsection{The uncertainty of the classical limit line}
In the third model, the extractable quantum randomness in the raw data is determined by the single photon start-stop event count rate under the classical limit line~\cite{photon_anti_re_sup,loudon2000quantum_sup,fox_book_2006_sup}
\begin{equation}
\begin{split}
r_{\mathrm{rand}}&=r_{\mathrm{A}}\sqrt{1-g^{(2)}_{\mathrm{fit}}(0)}\times r_{\mathrm{B}}\sqrt{1-g^{(2)}_{\mathrm{fit}}(0)}\times \int_{-t}^{t} g^{(2)}_{\mathrm{fit}}(\tau)d\tau\\
&=(1-g^{(2)}_{\mathrm{fit}}(0))\times(\eta_{\mathrm{A}}T I_{\mathrm{in}}-\frac{(\eta_{\mathrm{A}}T I_{\mathrm{in}})^2\int_0^{\tau_{\mathrm{dead}}^{\mathrm{A}}}g^{(2)}_{\mathrm{fit}}(\tau)d\tau}{4})(\eta_{\mathrm{B}}R I_{\mathrm{in}}-\frac{(\eta_{\mathrm{B}}R I_{\mathrm{in}})^2\int_0^{\tau_{\mathrm{dead}}^{\mathrm{B}}}g^{(2)}_{\mathrm{fit}}(\tau)d\tau}{4})\\&\quad \times\int_{-t}^{t} g^{(2)}_{\mathrm{fit}}(\tau)d\tau\\
&=(1-g^{(2)}_{\mathrm{fit}}(0))\times(\eta_{\mathrm{A}}T -\frac{(\eta_{\mathrm{A}}T)^2 I_{\mathrm{in}}\int_0^{\tau_{\mathrm{dead}}^{\mathrm{A}}}g^{(2)}_{\mathrm{fit}}(\tau)d\tau}{4})(\eta_{\mathrm{B}}R -\frac{(\eta_{\mathrm{B}}R)^2 I_{\mathrm{in}}\int_0^{\tau_{\mathrm{dead}}^{\mathrm{B}}}g^{(2)}_{\mathrm{fit}}(\tau)d\tau}{4})\\&\quad \times I_{\mathrm{in}}^2\int_{-t}^{t} g^{(2)}_{\mathrm{fit}}(\tau)d\tau~.
\end{split}
\label{r_rand}
\end{equation}

where $t$ satisfies $g^{(2)}_{\mathrm{fit}}(t)=1$. And the quantum fraction of the raw bits is defined as $r_{\mathrm{rand}}/r_{\mathrm{total}}$, then $p_c$, the classical noise probability, is $p_c=1-r_{\mathrm{rand}}/r_{\mathrm{total}}$. Under the classical limit line, the events 0-1 is taken as random bit \dzero, and 1-0 is taken as \done, the conditional min-entropy in this case is defined as
\begin{equation}
\begin{split}
H_{\infty}(\textbf{X}|\textbf{Y})&=-\mathrm{log}_2 \Big(p_c+(1-p_c)\big(\max\{p(\dzero\dzero),p(\dzero\done)\}+\max\{p(\done\dzero),p(\done\done)\}\big)\Big)\\
&=-\mathrm{log}_2 \Big(p_c+(1-p_c)f(\textbf{p})\Big)~.
\end{split}
\label{eqn:H_inf3}
\end{equation}

where $f(\textbf{p})=\mathrm{max}\{p(\dzero),p(\done),2p(\dzero\done),1-2p(\dzero\done)\}$.\\

For the convenience of description, without losing generality, we associate event pair ``AB" to random bit \dzero, and ``BA" to \done. For probabilities $p(\dzero)$ and $p(\done)$, there are two different situations. The first situation: when $t$ is larger than the half of the dead-time of the detectors, we have
\begin{equation}
p(\dzero)=p(AB)=p(A)\eta_{\mathrm{B}}R\big(\int_0^{t}g^{(2)}_{\mathrm{fit}}(\tau)d\tau-\eta_{\mathrm{B}}R\int_0^{\frac{\tau_{\mathrm{dead}}^{\mathrm{B}}}{2}}g^{(2)}_{\mathrm{fit}}(\tau)d\tau\times\underbrace{\int_0^{\frac{\tau_{\mathrm{dead}}^{\mathrm{B}}}{2}}g^{(2)}_{\mathrm{fit}}(\tau)d\tau}_{\text{incident photon within $\tau_{\mathrm{dead}}^{\mathrm{B}}/2$.}}\big)~.
\label{eqn:newp0p1}
\end{equation}

The probability here is very similar to~\ref{eqn:p(1|0)}, except we only consider short-time related photon events in this situation, so we replace `1' in~\ref{eqn:p(1|0)} with $\int_0^{t}g^{(2)}_{\mathrm{fit}}(\tau)d\tau$ here, where $p(A)$ is in~\ref{eqn:p(0)}. The other situation is when $t$ is smaller than the half of the dead-time of each detector, we need to change the formula inside the underbrace to $\int_0^{t}g^{(2)}_{\mathrm{fit}}(\tau)d\tau$, then 
\begin{equation}
p(\dzero)=p(AB)=p(A)\eta_{\mathrm{B}}R\int_0^{t}g^{(2)}_{\mathrm{fit}}(\tau)d\tau\big(1-\eta_{\mathrm{B}}R\int_0^{\frac{\tau_{\mathrm{dead}}^{\mathrm{B}}}{2}}g^{(2)}_{\mathrm{fit}}(\tau)d\tau\big)~.
\label{eqn:newp0p1_2}
\end{equation}

The equation of $p(\done)$ can be deduced in a similar way.\\

For the photon events under the classical limit line, the events pair \dzero~or \done~is are much less correlated than previous models, they can be treated as independent events, so we have
\begin{equation}
\begin{split}
p(\dzero\dzero)=p(\dzero)p(\dzero)~,\\
p(\dzero\done)=p(\dzero)p(\done)~,\\
p(\done\dzero)=p(\done)p(\dzero)~,\\
p(\done\done)=p(\done)p(\done)~.
\end{split}
\label{eqn:p(00)dzerodzero}
\end{equation}

Next we calculate the conservative conditional min-entropy in this model, similar to the second model, we have
\begin{equation}
H_{\infty}(\textbf{X}|\textbf{Y})=-\mathrm{log}_2 \big(p_{cq}+\Delta_{p_{cq}}\big)~.
\label{eqn:con_H_inf_3}
\end{equation}

where $p_{cq}=p_c+(1-p_c)f(\textbf{p})$, then $\Delta_{p_{cq}}=\sqrt{(\frac{\partial p_{cq}}{\partial p_c}\Delta_{p_c})^2+(\frac{\partial p_{cq}}{\partial f(\textbf{p})}\Delta_{f(\textbf{p})})^2}$. From $p_c=1-r_{\mathrm{rand}}/r_{\mathrm{total}}$, we get
\begin{equation}
p_c=1-\frac{(1-g^{(2)}_{\mathrm{fit}}(0))\times r_{\mathrm{A}}\times r_{\mathrm{B}}\times \int_{-t}^{t} g^{(2)}_{\mathrm{fit}}(\tau)d\tau}{r_{\mathrm{A}}+r_{\mathrm{B}}}~.
\label{eqn:pc}
\end{equation}

and the equation for $f(\textbf{p})$ is in~\ref{eqn:newp0p1_2},~\ref{eqn:newp0p1} and~\ref{eqn:p(00)dzerodzero}. From the equations of $p_c$ and $f(\textbf{p})$, we can see that they are dependent on some same parameters, including the dead-time of the two detectors, the detection efficiencies, and the beam-splitter ratio etc. This means that they are not independent from each other, so $\Delta_{p_{cq}}\neq\sqrt{(\frac{\partial p_{cq}}{\partial p_c}\Delta_{p_c})^2+(\frac{\partial p_{cq}}{\partial f(\textbf{p})}\Delta_{f(\textbf{p})})^2}$, $\Delta_{p_{cq}}$ should be derived directly from the experimental parameters
\begin{equation}
\begin{split}
\Delta_{p_{cq}}=\bigg((\frac{\partial p_c}{\partial T}\delta_T)^2+(\frac{\partial p_c}{\partial \eta_{\mathrm{A}}}\delta_{\eta_{\mathrm{A}}})^2+(\frac{\partial p_c}{\partial \eta_{\mathrm{B}}}\delta_{\eta_{\mathrm{B}}})^2+(\frac{\partial p_c}{\partial \tau_{\mathrm{dead}}^{A}}\delta_{\tau_{\mathrm{dead}}^{A}})^2+\\(\frac{\partial p_c}{\partial \tau_{\mathrm{dead}}^{B}}\delta_{\tau_{\mathrm{dead}}^{B}})^2+(\frac{\partial p_c}{\partial I_{\mathrm{in}}}\delta_{I_{\mathrm{in}}})^2+\frac{\partial p_c}{\partial t}\delta_{t})^2\bigg)^{\frac12}~.
\end{split}
\label{eqn:delta_pcq}
\end{equation}

where $t$ satisfies $g^{(2)}_{\mathrm{fit}}(t)=1$. Next we derive $\delta_{t}$. In our case, $\delta_{t}$ is characterized by the classical limit line. The classical limit line is determined by the normalization factor of the experimental anti-bunching curve. The normalization factor $N_{\mathrm{norm}}$ is calculated by

\begin{equation}\nonumber
N_{\mathrm{norm}}=r_{\mathrm{A}}  r_{\mathrm{B}} \tau_{\mathrm{rs}} T_{\mathrm{total}}
\end{equation}
where $\tau_{\mathrm{rs}}$ is the timing resolution of the start-stop event, $T_{\mathrm{total}}$ is the total integration time (the running time of the experiment). $N_{\mathrm{norm}}$ can be determined by multiple parameters, including the detection efficiency and dead-time of each detector, and the reflection and transmission coefficients. According to the propagation of uncertainty, we get the uncertainty of $N_{\mathrm{norm}}$

\begin{equation}\nonumber
\begin{split}
\Delta_{\mathrm{norm}}=\bigg((\frac{\partial N_{\mathrm{norm}}}{\partial R}\delta_R)^2+(\frac{\partial N_{\mathrm{norm}}}{\partial \eta_{\mathrm{A}}}\delta_{\eta_{\mathrm{A}}})^2+\\(\frac{\partial N_{\mathrm{norm}}}{\partial \eta_{\mathrm{B}}}\delta_{\eta_{\mathrm{B}}})^2+
(\frac{\partial N_{\mathrm{norm}}}{\partial I_{\mathrm{in}}}\delta_{I_{\mathrm{in}}})^2+\\(\frac{\partial N_{\mathrm{norm}}}{\partial \tau_{\mathrm{dead}}^{\mathrm{A}}}\delta_{\tau_{\mathrm{dead}}^{\mathrm{A}}})^2+(\frac{\partial N_{\mathrm{norm}}}{\partial \tau_{\mathrm{dead}}^{\mathrm{B}}}\delta_{\tau_{\mathrm{dead}}^{\mathrm{B}}})^2\bigg)^{\frac12}~.
\end{split}
\end{equation}

\noindent
Then the uncertainty of the classical limit line (i.e.\ $g^{(2)}_{\mathrm{fit}}(\tau)=1$) amounts to
\begin{equation}
\Delta_{1}=1\times\frac{\Delta_{\mathrm{norm}}}{N_{\mathrm{norm}}}~.
\label{eqn:delta_unit_line}
\end{equation}

and the uncertainty of the background line $g^{(2)}_{\mathrm{fit}}(0)$ is

\begin{equation}
\Delta_{g^{(2)}_{\mathrm{fit}}(0)}=g^{(2)}_{\mathrm{fit}}(0)\frac{\Delta_{\mathrm{norm}}}{N_{\mathrm{norm}}}~.
\label{eqn:g2_fluctuation}
\end{equation}

From the uncertainty of classical limit line, $\delta_{t}=t-t'$ can be deduced, where $t'$ satisfies the equation $g^{(2)}_{\mathrm{fit}}(t')=1-\Delta_{1}$. Then $\Delta_{p_{cq}}$ can be derived.

\subsection{Two-universal hashing and the 11.5$\sigma$ error bound}
Two-universal hashing can be used as a randomness extractor~\cite{frauchiger_a_2013_sup}. The main idea of this extractor is explained in~\cite{troyer_ir_2012_sup}: For a $n$ bits random string $X$, we want to extract a random string $Y$, which is $k$ bits long, $k<n$. The quality of a random extractor is quantified by the probability that the output random string $Y$ deviates from a perfect uniformly distributed $k$-bit string.\\

Two-universal hashing can be simply done by a bit-matrix-vector multiplication with a seed $m$ (a random matrix with dimension $n\times k$)
\begin{equation}\nonumber
y_i=\sum_{j=1}^n m_{i,j}x_{j}~.
\end{equation}
where $y_i\in Y$, and $x_j\in X$.

Let $\epsilon$ be the deviation, and $H$ be the entropy in the input string $X$, then with two-universal hashing, the deviation $\epsilon$ is bound by~\cite{tomamichel_ieee_2011_sup,frauchiger_a_2013_sup}
\begin{equation}\nonumber
\epsilon=2^{-(Hn-k)/2}~.
\end{equation}

In our case, the error bound $\epsilon$ is decided to be $2^{-100}$ or $10^{-30}$~\cite{troyer_ir_2012_sup}. The physical meaning of this bound is that in the age of the universe, even when one million different random strings are hashed by this two-universal hashing, we cannot observe any deviation from a perfect uniform randomness in the output string.\\

The $11.5\sigma$ error bound mentioned below can be derived by the following Gaussian distribution
\begin{equation}\nonumber
1-\int_{-m\sigma}^{m\sigma}\frac{1}{\sqrt{2\pi\sigma^2}}e^{-\frac{x^2}{2\sigma^2}}dx=2^{-100}~.
\end{equation}
the solution of $m$ is $m=\sqrt{2} \mathrm{erf}^{-1}\Big(\frac{2^{100}-1}{2^{100}}\Big)\approx11.5$, where $\mathrm{erf}$ is the Error Function.

\section{Experimental results}
In this section, some calculation details of the experiment results are described. First we estimate the 1$\sigma$ error bound of the experimental devices. Reasonably, we have the following assumption: the error bound of the beam-splitter is $\delta_R=0.004$, the error bound of the two detectors' detection efficiencies is $\delta_{\eta_{\mathrm{A}}}=\delta_{\eta_{\mathrm{B}}}=0.01$, the error bound of dead-time is $\delta_{\tau_{\mathrm{dead}}^{A}}=\delta_{\tau_{\mathrm{dead}}^{B}}=10~\mathrm{ns}$, and the error bound of the incident photon rate is $\delta_{I_{\mathrm{in}}}=\sqrt{I_{\mathrm{in}}}$.\\ 

The values of the experimental parameters are estimated as

\begin{table}[h]
\begin{center}
\begin{tabular}{ |l|*{7}{c|} }
\hline
 $\tau_{\mathrm{dead}}^{\mathrm{A}}$& $\eta_{\mathrm{A}}$ & T &$\tau_{\mathrm{dead}}^{\mathrm{B}}$& $\eta_{\mathrm{B}}$ & R & $I_{\mathrm{in}}$ \\
\hline
43.5ns & 60\%& 0.39 &  42.9ns & 60\% &0.61 & 166~kcps\\
\hline
\end{tabular}
\caption{\textbf{The parameter value of experimental devices}}
\label{table:t2}
\end{center}
\end{table}

For the first model, according to~\ref{eqn:p(0)} and~\ref{eqn:delta_p(1|0)}, $p(1)=0.6100$, and $p(10)=0.2379$, considering the different conditions in~\ref{eqn:H_inf conditions}, the output conditional min-entropy per raw bit is 0.7132 bits, when introducing the 11.5$\sigma$ error bound, from~\ref{eqn:con_H_inf_1}, the conservative conditional min-entropy is 
\begin{equation}\nonumber
H_{\infty}(X|Y)=-\mathrm{log}_2\big(p(1)+11.5\Delta_{p(1)}\big)=0.5559~.
\end{equation}

This means that the conservative randomness per raw bit is 0.5559 bits, where $\Delta_{p(1)}=0.006$ is from ~\ref{eqn:delta_p(0)}. For this model, the experimental data gives us a value of $H_{\infty}(X|Y)=0.7128$, which is within the given error bound. \\

With the conservative 11.5$\sigma$ error bound, by two-universal hashing, approximate $3.10\times10^{10}$ unbiased random bits can be extracted from the raw random bits, and the unbiased random number generation speed is $3.10\times10^{10}\mathrm{bits}/608125\mathrm{s}=5.1\times10^{4}$ bits per second.\\

In the second model, by limiting the raw random data to single photon events, we exclude the uncorrelated background noise from the total photon events. According to this model, we get $p(1)=0.6100$, $g^{(2)}_{\mathrm{fit}}(0)=0.1116$, and the fraction of the uncorrelated background noise $p_e=1-\sqrt{1-0.1116}=0.0575$. From~\ref{eqn:H_inf 2}, the output quantum randomness per raw bit is 0.6612 bits. From~\ref{eqn:con_H_inf_2}, considering the 11.5$\sigma$ error bound, the conservative conditional min-entropy is
\begin{equation}\nonumber
H_{\infty}(X|Y)=-\mathrm{log}_2 \Big(p_{eq}+11.5\Delta_{p_{eq}}\Big)=0.5168~.
\end{equation}

where $p_{eq}=0.6324, \Delta_{p_{eq}}=0.0058$ can be calculated from~\ref{subsec:error_bound}. In the experimental data, $g^{(2)}_{\mathrm{fit}}(0)=0.1114$, and $p(1)=0.6101$, using~\ref{eqn:H_inf 2}, the extractable quantum randomness per raw bit is 0.6610 bits, which is covered by the 11.5$\sigma$ error bound. Then with the strict 11.5$\sigma$ error bound, the total quantum random number in the raw data is $2.88\times10^{10}$ bits, and the generation speed is about $2.88\times10^{10}\mathrm{bits}/608125\mathrm{s}=4.74\times10^{4}$ bits per second.\\

In the third model, a more strict quantum randomness extraction method is introduced. In most cases, when $g^{(2)}_{\mathrm{fit}}(0)<1$ the photons are non-classical~\cite{photon_anti_re_sup,loudon2000quantum_sup,fox_book_2006_sup}, and it is assumed that only single photons could reach the domain of $g^{(2)}_{\mathrm{fit}}(0)<1$. To limit our photon events to single photons, we only consider events that are below the classical limit line and above the background noise line.\\

From~\ref{eqn:pc}, $p_c=0.99939$. Solving $g^{(2)}_{\mathrm{fit}}(t)=1$, we have $t=21.13~\mathrm{ns}$, since it is smaller than the half of the dead-time of two detectors, according to~\ref{eqn:newp0p1_2}, we have $p(\dzero)=0.500051$ and $p(\done)=0.499949$. From~\ref{eqn:p(00)dzerodzero} and~\ref{eqn:H_inf3}, the output quantum randomness per raw bit is $H_{\infty}(\textbf{X}|\textbf{Y})=4.396\times10^{-4}$ bits. Considering the 11.5$\sigma$ error bound, from~\ref{eqn:con_H_inf_3} and~\ref{eqn:delta_pcq}, we compute the extractable quantum randomness per raw bit is
\begin{equation}\nonumber
H_{\infty}(\textbf{X}|\textbf{Y})=-\mathrm{log}_2 \Big(p_{cq}+11.5\Delta_{p_{cq}}\Big)=3.746\times10^{-4}~.
\end{equation}

In the experimental data, the data below the classical limit line is short-time related. By post-processing, the total events below the classical limit line is 32607956 bits, these bits are generated in a time without refocusing, which is 558000 seconds, then the tuple events per second is about 58.43 bits. The generation speed for the raw bits is about~100~kcps, thus $58.43/100000\approx5.843\times 10^{-4}$ is the fraction of quantum random bits in the raw bits, the rest is considered as classical noise, $p_c=1-5.843\times10^{-4}=0.999416$. In these quantum random bits, we have 16305447 bits which are \dzero, and 16302509 bits \done, $p(\dzero)=0.500045, p(\done)=0.499955$. Then the extractable quantum randomness from per raw bit is $4.211\times10^{-4}$, which does not exceed the 11.5$\sigma$ error bound.\\

With the given error bound, the number of quantum random bits in the raw data is $2.09\times10^{7}$ bits, the generation speed of these quantum random bits is $2.09\times10^{7}\mathrm{bits}/608125\mathrm{s}=$34.37 bits per second. 


\end{document}